%% file: acml22.tex
\documentclass[wcp]{jmlr}

\usepackage{xcolor}

\usepackage{longtable}
\usepackage{algorithm}  
\usepackage[noend]{algpseudocode}
\usepackage{bbding}
\usepackage{enumitem}
\usepackage{dashrule}
\usepackage{graphicx}
\usepackage{soul}
\usepackage[textsize=footnotesize,textwidth=2.5cm]{todonotes}

\usepackage{booktabs}

\makeatletter
\let\Ginclude@graphics\@org@Ginclude@graphics 
\makeatother

\jmlrvolume{}
\jmlryear{2022}

\title[FLVoogd]{FLVoogd: Robust And Privacy Preserving Federated Learning}

\author{\Name{Yuhang Tian} \Email{y.tian-13@student.tudelft.nl}\\
  \addr Delft University of Technology
  \AND
  \Name{Rui Wang} \Email{r.wang-8@tudelft.nl}\\
  \addr Delft University of Technology
  \AND
  \Name{Yanqi Qiao} \Email{y.qiao@tudelft.nl}\\
  \addr Delft University of Technology
  \AND
  \Name{Emmanouil Panaousis} \Email{e.panaousis@greenwich.ac.uk}\\
   \addr University of Greenwich
   \AND
  \Name{Kaitai Liang} \Email{kaitai.liang@tudelft.nl}\\
   \addr Delft University of Technology
 }

\begin{document}
\frenchspacing
\maketitle

\begin{abstract}
In this work, we propose FLVoogd, an updated federated learning method in which servers and clients collaboratively eliminate Byzantine attacks while preserving privacy. In particular, servers use automatic  {Density-based Spatial Clustering of Applications with Noise (DBSCAN)}
combined with S2PC to cluster the benign majority without acquiring sensitive personal information. Meanwhile, clients build dual models and perform test-based distance controlling to adjust their local models toward the global one to achieve personalizing.  {Our framework is automatic and adaptive that servers/clients don't need to tune the parameters during the training. In addition, our framework leverages Secure Multi-party Computation (SMPC) operations, including multiplications, additions, and comparison, where costly operations, like division and square root, are not required.} Evaluations are carried out on some conventional datasets from the image classification field. The result shows that FLVoogd can effectively reject malicious uploads in most scenarios; meanwhile, it avoids data leakage from the server-side. 
\end{abstract}
\begin{keywords}
federated learning; secure-multi-party computation; differential privacy 
\end{keywords}

\input{sections/introduction}

\input{sections/background}

\input{sections/design}

\input{sections/experiment}

\input{sections/conclusion}
\setlength{\bibsep}{0em}
\bibliography{acml22}
\end{document}

%% file: sections/introduction.tex
\section{Introduction}
{Unlike the centralized learning setting, where a server collects substantial users' data to build a  model for predictions, Federated Learning (FL) requires the model parameters that clients train independently with their data and devices. In FL paradigm frameworks, such as FedAvg~\cite{FedAvg}, the server iteratively aggregates local models trained by individuals and sends the global one back to clients for their local updates, to achieve collaborative training. Since no actual data is sent to the server, this paradigm was considered privacy-preserving.} 
In the past half-decade, FL has been widely researched and applied in many fields such as image recognition~\cite{li2021model}, Natural Language Processing (NLP)~\cite{liu2021federated}, financial system~\cite{long2020federated}, and medical care~\cite{dayan2021federated}.

However, such a framework still faces two main challenges - privacy and security. 
On the one hand, the conventional FL setting cannot get rid of the disclosure of clients' information and even enlarge the attacking surface~\cite{aono2017privacy}.
Not only can the server be an adversary to infer the information from local models sent by clients in this scenario, but also every participant can perform the inference attack on the global model constructed by each individual. {Therefore, some research employs SMPC to encrypt the uploads, such as~\cite{FLGuard}. However, the cost of the design or operations is high, especially when dealing with the division and the square root.}
On the other hand, without countermeasures, adversaries can arbitrarily substitute the data with the poisoned one~\cite{attackTails} or even directly change the upload into a meaningless random number, leading their uploads to betray the regulation. {To eliminate the attacking consequence, some research, like~\cite{ditto} and~\cite{DeepSight}, builds a practical defensive framework, but with too unintuitive hyper-parameters to tune for different situations.}

{To improve the efficiency and save expensive operations}, we develop an updated framework that combines SMPC~\cite{crypten, sympc}, DBSCAN~\cite{2-dbscan}, differential privacy~\cite{DPFLclientlevel}, and  {personalized local model}~\cite{ditto}
to eliminate the malicious uploaded parameters without revealing any sensitive information and guarantee $(\epsilon,\delta)$-DP after the aggregation. 
 {Compared with the past research, our framework 1) filters the abnormal uploads without knowing their sensitive information; 2) performs the training process adaptively, requiring no parameter tuning; 3) uses SMPC operations efficiently supported by most protocols.}
We leverage the conventional image classification dataset to evaluate the framework. The results show that the filter can reject the Byzantine attacks under most situations without degrading the model performance, and the trade-off between predicting accuracy and DP strength can be customized for different scenarios. 

%% file: sections/background.tex
\section{Background and Problem Setting}
\subsection{Adversarial Attack}
\label{subsec: adversarial_attack}
Six different state-of-art attacking schemes are used to test our FL defenses' robustness and named from A1 to A6.\\
$\bullet$\textbf{Random upload} (A1): As its name suggests, the adversary substitutes the factual update with a random noise chosen from $X\sim\mathcal{N}(0,1)$. Consequently, the average of parameters can arbitrarily deviate from $w_{avg} = \frac{1}{n}\sum_{i=1}^nw_i$ to
${w_{dev}} = \frac{1}{m}\sum_{i=1}^mw_i+\frac{1}{n-m}\sum_{i=m+1}^n\mathcal{N}(0,1)$, where $m$ is the number of honest updates and $(n-m)$ is the number of malicious updates.\\
$\bullet$\textbf{Krum attack} (A2): It is designed  to crack the Krum aggregation rule. In a nutshell, Krum selects one vector from a set of $n$ vectors that is the most comparable to the rest. Even if a compromised client gives the chosen vector, the impact is limited in this situation. However, when adversaries try to invalidate the Krum aggregation rule, they can conspire to elaborate a set of vectors to make $KR(w_1',...,w_f',...,w_n)$ output $w_1'$ such that $w_1'$ mostly inversely differs from the true selected one without being attacked~\cite{3-poisonattack}.\\
$\bullet$\textbf{Trimmed-mean attack} (A3): Trimmed-mean sorts $n$ updates for each $j^{th}$ parameter $sort(w_{1j},...,w_{nj})$, eliminates the highest and smallest $\beta$ amount from the sorted list, and averages the remaining $(n-2\beta)$ parameters as the global model's $j^{th}$ parameter~\cite{3-trimmedmean}. To enervate this aggregation rule, adversaries collude to submit deviating models in the opposite direction that the global model would change in the absence of attacks.\\
$\bullet$\textbf{Label flipping} (A4): Each adversary converts the label of a sample from $l$ to $L-l-1$, where $l$ is the truth label of the sample, and $L$ is the total number of classes~\cite{3-labelflipping}. For instance, adversaries label digit ``0'' as ``9'' and digit ``9'' as ``0'' to label-flip the MNIST data.\\
$\bullet$\textbf{Backdoor triggering} (A5): This kind of attack is also known as trojan attacks~\cite{3-trojan}. The adversary inserts a specific pattern into training samples or uses existing ones to render the corresponding testing samples with that pattern classified as the desired class. This pattern functions as a trigger. After the global model learns this pattern, it will be triggered and output the misled prediction. If the adversary uses the existing pattern in the sample, this backdoor attack is a semantic backdoor attack~\cite{DeepSight}.\\
$\bullet$\textbf{Edge-case attack}(A6): Under the edge-case attack setting, adversaries aim to attack the heavy-tail of the prediction~\cite{attackTails}. They try to find or manufacture samples that the model predicts correctly but with a comparably low confidence value; then, they label those samples with a label they want. The intuition behind it is that the model cannot assure the correctness of predictions even if the result is correct, as the predicting score is not such high, so it can be easily misled by the attacker who feeds those edge-case samples with wrong labels.

\subsection{DBSCAN}
DBSCAN is initially designed for clustering and distinguishing the noise from the high dimensional database depending on the variance of density~\cite{2-dbscan}. A non-negligible quantity of samples should be in the cluster if a cluster is formed, while the cluster can hardly be formed in areas where samples are located sparsely. These 
``depopulated zones'' can be used as gaps to separate the different classes and to sift out noisy samples. We will consistently follow some of the concepts and symbols used in~\cite{2-dbscan}. $N_{Eps}(p)$ represents neighbors of a point $p$ within a range with radius $Eps$ ($Eps$ is a preset hyper-parameter). A point $p$ is a $core point$, if $|N_{Eps}|\geq{MinPts}$ ($MinPts$ is a preset hyper-parameter). In addition, a $core point$ is the centroid of a cluster, so in other words, a cluster is only formed when its centroid is a $core point$. A point $p$ is a $border point$, if its neighbours contain at least one $core point$. It should be noted that a point can be a $border point$ for different clusters, but it will be only assigned to a unique cluster eventually, and it depends on which cluster it assigns the point to first. If a point is neither a $core point$ nor $border point$, it will be classified as noise.

\subsection{SMPC}
As mentioned, uploading weights instead of the raw data to the server is not privacy-preserving. As shown in~\cite{GANinference}, model parameters can disclose some information about individual data. For example, adversaries can use Generative Adversary Networks (GANs) to reconstruct the class representatives from the aggregated parameters. This powerful reconstruction is more harmful if it happens on the server side because the server can steal the class representatives from each individual uploading. To avoid revealing the uploads to the server, SMPC can be used for private aggregation, and the result will only be revealed eventually. Following the structures in~\cite{flame, FLGuard, DeepSight}, we will use Secure 2-party Computation (S2PC), a ramification of SMPC, to guarantee that the individual upload will not be plain-text to the server. Under the S2PC setting, each client will not directly send the model parameters to the server but separate the upload into two parts and share one with the server for aggregation and another with the external server. As both servers hold merely one piece of the secret,  the secret cannot be known if they do not collude. Based on the secret sharing scheme, each server can do arithmetic operations relying on its own share and through some communication. To achieve this target, two libraries CrypTen\footnote{\href{https://github.com/facebookresearch/CrypTen}{https://github.com/facebookresearch/CrypTen}}~\cite{crypten} and SyMPC\footnote{\href{https://github.com/OpenMined/SyMPC}{https://github.com/OpenMined/SyMPC}}~\cite{sympc} derived from PySyft are used for the experiment. Both of them use secret sharing but with different protocols to achieve S2PC. CrypTen is currently designed only for semi-honest parties, while SyMPC can tolerate minor malicious parties.

\subsection{Security Assumption}
{
We primarily consider possible server-side and client-side risks.
There are two types of servers in our setting. 
Firstly, servers can be honest-but-curious who infer the actual data or relevant information from uploads while heeding the regulation.
Secondly, if FLVoodg runs under SyMPC-Falcon~\cite{wagh2020falcon}, servers can be malicious (minority) who betray the secure aggregation rule and send an incorrect model back to participants. 
In terms of participants, in each round, less than half of them can be malicious and perform byzantine attacks~\ref{subsec: adversarial_attack} to deteriorate the performance of the global model. 
{In addition, any client can be curious about information from others and performs client-level inference attacks~\cite{DPFLclientlevel}, inferring whether a particular client participates in the training, given a specific dataset of that client.}
}

%% file: sections/design.tex
\section{FLVoogd Overview and Design}
\subsection{FLVoogd Server}
\begin{figure}[htb]
    \centering
    \includegraphics[width=1.0\textwidth]{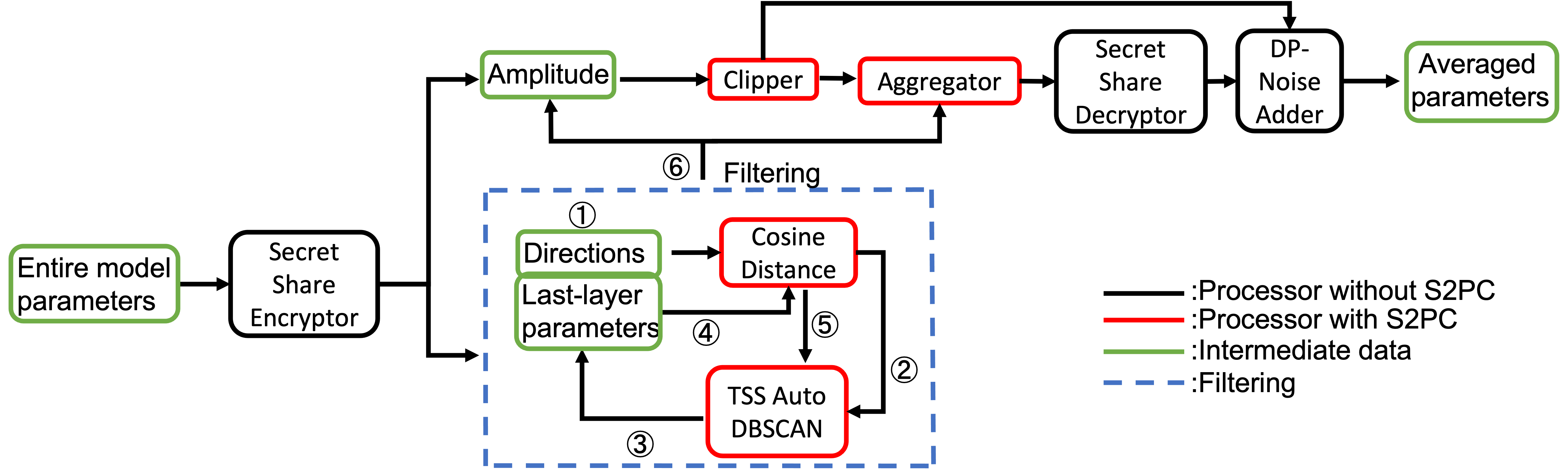}
    \caption{FLVoogd server framework.}
    \label{fig: flvoogd_enc}
\end{figure}

The overview of our framework, FLVoogd with SMPC, is shown in Fig.~\ref{fig: flvoogd_enc}. Initially, each client computes the $l2$-norm of its uploaded parameters $w_i$ as the amplitude $||w||_i\gets\sqrt{\sum_jw_{ij}^2}$ and unitizes the parameters to $\overline{w}_i\gets\frac{w_i}{||w||_i}$ as the direction. The unitizing can simplify the latter SMPC computations, e.g., cosine distance where division and square root operations are saved. Besides, the client performs the same for the parameters from the last layer and obtains the unitized last layer parameters $\overline{v}_i$. The reason for extracting the last layer is that parameters in the final layer reveal more explicit information relevant to the dataset's distribution~\cite{DeepSight} which can be used for distinguishing backdoor uploads. Then, the client secretly shares the direction $\overline{w}_i$, the amplitude $||w||_i$, and last-layer direction $||v||_i$ with two parties, the server for aggregation  and external server. If the SMPC protocol uses Falcon, one more server is added and receives the third share from the client.

Once servers receive the uploads from the selected clients, they can perform secure aggregation. It is supposed that the number of received uploads is $n$, and the clipping boundary for the current round is $c^{(t)}$. The server firstly carries out the filtering process, the block with a light yellow background in Fig.~\ref{fig: flvoogd_enc}, and the procedure can be divided into six steps. \textcircled{
1}: In the first step, the server uses the directional vector $\overline{w}$ to compute the cosine distance matrix $M_{cos}$ by Eq.~\eqref{eq: cosinedist}, {where $M_{cosij}=cos-dist(i,j)$ for $i\neq{j}$, $M_{cosij}=0$ for $i=j$, $i$ or $j$ denotes the client's index, and $u$ denotes the parameter's index.} Since the vectors are unit vectors, the cosine distance between two vectors can be simplified into a dot production. The computation is collaboratively completed by two/three servers, involving the addition and multiplication among secret shares. \textcircled{$2$}: In the second step, the server feeds the Total-Sum-of-Square (TSS)-based DBSCAN with the distance matrix from the former step. DBSCAN calculates the TSS by Eq.~\eqref{eq: tssdist} for each pair of rows to obtain a new distance matrix $M_{tss}$. {$M_{cos}$ provides the directional similarity, while $M_{tss}$ enlarges the variance of counter-directions and narrows the variance of identical directions such that the filter can capture the difference more easily.} There are two hyper-parameters for DBSCAN, $Eps$ and $MinPts$. As honest-majority is the basic assumption, $MinPts$ is set to $\lfloor n/2\rfloor+1$ and $Eps$ is the average of the median ($\lfloor n/2\rfloor$) in each row of the distance matrix $M_{tss}\in{\mathbb{R}^{n\times{n}}}$. This setting for $Eps$ and $MinPts$ guarantees the DBSCAN can automatically adjust its radius accordingly through the whole process without manual involvement, and the selection makes full use of the honest-majority assumption. This step includes the addition, multiplication, and comparison of shares. \textcircled{3}: In the third step, DBSCAN filters out the noise and minority group and returns indices of the majority group. The server selects the corresponding clients' parameters from the last layer according to the indices. \textcircled{$4$}: For the next step, similar to step 1, the server again computes the cosine distance matrix but now uses the last-layer parameters. Then, the server obtains cosine distance matrix $M'_{cos}$. \textcircled{$5$}: The fifth step is identical to the second step. \textcircled{$6$}: In the final steps, DBSCAN outputs the indices that the server will consider as benign.
\begin{align}
    \label{eq: cosinedist}
    cos-dist(i,j) 
    &= \frac{\sum_u\overline{w}_{i,u}\overline{w}_{j,u}}{\sqrt{\sum_u\overline{w}_{i,u}^2}\sqrt{\sum_u\overline{w}_{j,u}^2}}
    = \sum_u\overline{w}_{i,u}\overline{w}_{j,u}
    = \overline{w}_{i}\cdot\overline{w}_{j} \\
    \label{eq: tssdist}
    tss-dist(i,j) 
    &= \sum_u{(M_{cosi,u}-M_{cosj,u})^2}
\end{align}

After knowing which clients are considered as benign ones, the server clips their amplitudes before performs the aggregation by $||w||_i=\min(||w||_i, c^{(t)})$. On the one hand, it restrains the abnormally large amplitude and controls the next descent step size. On the other hand, it provides the $l2$-sensitivity for the DP budget tracer. Furthermore, during the clipping, the clipper records the ratio of clients not being clipped as $\hat{\gamma}$. The expected clipping ratio is set as $\gamma$. The next round clipping boundary $c^{(t+1)}$ is updated by Eq.~\eqref{eq: adaptiveclip} where $\eta_c$ is the learning rate of the clipper. If the actual non-clipping number of clients is larger than expected, the clipping boundary will decrease to cut more clients in the next round; otherwise, it will increase to be looser. The exponential base guarantees that any adjustment is a positive number. SMPC's operation in this step requires comparing a share with a public number. 

\begin{equation}
    \label{eq: adaptiveclip}
    c^{(t+1)} = c^{(t)}\cdot\exp(-\eta_c(\hat{\gamma}-\gamma))
\end{equation}

The aggregation rule is simply averaging benign clients' uploads by Eq.~\eqref{eq: aggregation}. It is supposed that the number of clients after filtering is $m (\leq{n})$. After obtaining the merged model $w_{global}$, servers collaboratively reveal and announce the plain text of $w_{global}$. The aggregation contains the addition and multiplication of shares, and the multiplication of shares and public numbers.

\begin{equation}
    \label{eq: aggregation}
    w_{global}=\frac{1}{m}\sum\limits_{i=1}^m \overline{w}_i\cdot\min(||w||_i,c^{(t)})
\end{equation}

The server eventually knows the global parameter till finishing aggregation, and local parameters are already merged into one; thus, the server has no idea of individual local updates. Before sending the global update back to clients, the server adds Gaussian noise to provide a differential privacy guarantee to {defend against client-level inference from the client side}. The mean is $0$, and the standard deviation is the maximum $l2$-sensitivity multiplied by a coefficient $\sigma$ that represents the strength of DP. Thanks to the clipping, all uploads are bounded into a sphere whose radius is exactly the clipping boundary $c^{(t)}$. Therefore, the noise is added following Eq.~\eqref{eq: clippingDPnoise}. Notably, the noise is added to the sum of updates not after averaging. DP-Noise Adder also tracks the DP budget for the server because it knows the number of clients used in this round and the amplitude of Gaussian noise. Finally, $||\tilde{w}_{global}||$ is compared with $||c^{(t)}||$. If $||\tilde{w}_{global}||>||c^{(t)}$, from which the server deduces that the amount of noise is added too much, the server will scale down $||\tilde{w}_{global}||$ to a smaller value by Eq.~\eqref{eq: postprocessing}. This operation follows the post-processing property of differential privacy so that $(\epsilon,\delta)$ cannot be influenced. This post-processing is equivalent to adjusting the model learning rate lower after knowing the noise influences too much on the result. The algorithm of FLVoogd server is manifested in Alg.~\ref{alg: flvoogd_enc_server}.

\begin{align}
    \label{eq: clippingDPnoise}
    \tilde{w}_{global} & = \frac{1}{m}\{\sum\limits_{i=1}^m \overline{w}_i\cdot\min(||w||_i,c^{(t)}) + \mathcal{N}(0,\sigma^2\cdot{(c^{(t)})^2})\} \\
    \label{eq: postprocessing}
    \tilde{w}_{global} & := \tilde{w}_{global}\cdot\min(1,\frac{c^{(t)}}{||\tilde{w}_{global}||})
\end{align}

\begin{algorithm}[htb]
    \caption{FLVoogd server algorithm}
	\label{alg: flvoogd_enc_server}
	\begin{algorithmic}[1]
	\State \textbf{Input}: 
	\State $\mathcal{C}$, $N$ \Comment{$\mathcal{C}$ is the set of clients, $N=|\mathcal{C}|$}
	\State $T$, $q$ \Comment{$T$ is the number of training iteration, $q$ is the sampling ratio} 
	\State $c^{(0)}$, $\gamma$, $\eta_c$  \Comment{$c^{(0)}$ is the initial clipping boundary, $\gamma$ is the expected clipping ratio, $\eta_c$ is Clipper's learning rate}
	\State $\sigma$, $\delta$ \Comment{$\sigma$ is the coefficient to control the noise strength, $\delta$ is for DP}
	\For{round $t$: $1,2,...,T$}
        \State $\mathcal{C}^{(t)}$, $n\gets$ subsample($\mathcal{C}$, $N$, $q$) \Comment{$n=|\mathcal{C}^{(t)}|$}
	    \For{$client_i\in\mathcal{C}^{(t)}$}
            \State $\overline{w}^{(t)}_i$, $||w||^{(t)}_i$, $\overline{v}^{(t)}_i\gets{client_i(t,send)}$ \Comment{$\overline{w}$ is the unit vector of weight difference, $||w||$ is the norm of weight difference, $\overline{v}$ is the unit vector of last layer's weight difference}
        \EndFor
        \State $idx^{(t)}_{f1}$, $n^{(t)'}\gets$ Auto\_DBSCAN($\{\overline{w}^{(t)}_1, \overline{w}^{(t)}_2,...,\overline{w}^{(t)}_n\}$) by Alg.~\ref{alg: autodbscan} \Comment{$n^{(t)'}=|idx^{(t)}_{f1}|$}
        \State $idx^{(t)}_{f2}$, $n^{(t)''}\gets$ Auto\_DBSCAN($\{\overline{v}^{(t)}_i:i\in{idx_{f1}}\}$) by Alg.~\ref{alg: autodbscan} \Comment{$n^{(t)''}=|idx^{(t)}_{f2}|$}
        \State $w_{global}^{(t)}\gets0$, $\hat{\gamma}^{(t)}\gets0$
        \For{index $i\in{idx^{(t)}_{f2}}$}
            \If{$||w||^{(t)}_i>c^{(t)}$}
                \State $||w||^{(t)}_i\gets{c^{(t)}}$
            \Else
                \State $\hat{\gamma}^{(t)}\gets{\hat{\gamma}^{(t)}+1}$
            \EndIf
            \State $w_{global}^{(t)}\gets{w_{global}^{(t)}+||w||^{(t)}_i\cdot\overline{w}^{(t)}_i}$
        \EndFor
        \State $w_{global}^{(t)}\gets\frac{w_{global}^{(t)}}{n^{(t)''}}$, $\hat{\gamma}^{(t)}\gets\frac{\hat{\gamma}^{(t)}}{n^{(t)''}}$
        \State $c^{(t+1)}\gets{c^{(t)}\cdot\exp(-\eta_c(\hat{\gamma}^{(t)}-\gamma))}$, $\epsilon^{(t)}\gets$ DP\_budget($\frac{n^{(t)''}}{N}$, $\sigma$, $\delta$)
        \State $\tilde{w}_{global}^{(t)}\gets{w_{global}^{(t)}+\frac{1}{n^{(t)''}}\mathcal{N}(0,\sigma^2(c^{(t)})^2)}$ \Comment{satisfying $(\epsilon,\delta)$-differential privacy}
        \State $\tilde{w}_{global}^{(t)}\gets\tilde{w}_{global}\cdot\min(1,\frac{c^{(t)}}{||\tilde{w}_{global}||})$ \Comment{satisfying post-processing}
        \State $client_i(t,receive)\gets\tilde{w}_{global}^{(t)}$
    \EndFor
	\end{algorithmic}  
\end{algorithm}

\begin{algorithm}[htb]
    \caption{Auto DBSCAN}
	\label{alg: autodbscan}
	\begin{algorithmic}[1]
	\State \textbf{Input}: $W$  \Comment{$W\in\mathbb{R}^{n\times{m}}$ represents $n\times{m}$ matrix where each row is a client's unit vector from $n$ clients and the dimension of the vector is $m$}
	\State \textbf{Output}: $idx$, $|idx|$ \Comment{$idx$ is a list of indices of benign clients}
	\State $M_{cos}\gets$ CosDist($W$) by Eq.~\eqref{eq: cosinedist}
	\State $M_{tss}\gets$ TSSDist($M_{cos}$) by Eq.~\eqref{eq: tssdist}
	\For{row $i$: $1,2...,n$}
        \State $median_i\gets$ quickMedian($M_{tss,i}$)
    \EndFor
    \State $median\gets\frac{1}{n}\sum_{i=1}^n{median_i}$
    \State $Eps\gets{median}$, $MinPts\gets{n//2+1}$ 
    \State $idx\gets{DBSCAN(M_{tss}, Eps, MinPts, precomputed)}$
    \State \Return $idx$, $|idx|$
	\end{algorithmic}  
\end{algorithm}

\subsection{FLVoogd Client}

Referencing the idea from Ditto~\cite{ditto}, each FLVoogd's client builds two identical models, namely, the global model and the local model. 
{Varying from~\cite{FLTrust, FLGuard, flame, DeepSight} where the server entirely takes the responsibility of a robust model, clients can share that responsibility locally. This defensive scheme builds a gap between the self-used local and global models to enhance the robustness and offers personalization to users.}
In each round, the client receives the averaged aggregated weight difference $\tilde{w}_{global}$ from the server and updates the weight of the global model accordingly by $W_{global} := W_{global} + \tilde{w}_{global}$. In contrast, the local model is not updated in this step. After updating the locally global model, the client tests the model accuracy using evaluation data and obtains the testing accuracy $acc_{ref}$. 

The client feeds the partial training data to the global and local models in each mini-batch iteration. It is supposed that there is a coefficient $\lambda_{ditto}$ to control the distance of the local model from the global model. The objective function of the local model becomes like Eq.~\eqref{eq: localobjectivefunction}, where $F(\cdot)$ is the objective function for the global model and originally for the local model. The change in the local model's objective function now is that the client adds an additional $l2$-regularization term to force the local model to approximate the global model. Consequently, the local model can learn from the global model, and the gap between them is constrained by $\lambda_{ditto}$.

\begin{equation}
    \label{eq: localobjectivefunction}
    \min\limits_{W_{local}} F'(W_{local};W_{global}) = F(W_{local}) + \frac{\lambda_{ditto}}{2}||W_{local}-W_{global}||^2
\end{equation}

Furthermore, Eq.~\eqref{eq: localobjectivefunction} can be converted into a gradient decent format shown in Eq.~\eqref{eq: dittogradientdecent}, where $\eta_{local}$ is the client's local learning rate. The formula shown in Eq.~\eqref{eq: dittogradientdecent} can be easily implemented by PyTorch where the client extracts the gradient and adds the $\lambda_{ditto}(W_{local}-W_{global}))$ term to it before running \textit{optimizer.step()}. 

\begin{equation}
    \label{eq: dittogradientdecent}
    g := g - \eta_{local}(\triangledown{F(W_{local})}+\lambda_{ditto}(W_{local}-W_{global}))
\end{equation}

Till now, the client has $\lambda_{ditto}$ as a controller to adjust the learning distance between the local and global models, but how to set an appropriate value $\lambda_{ditto}$ for the local model? Intuitively, if the global model is admirable and exemplary, we expect the local model to learn as much helpful information as possible from the global model; otherwise, we desire the local model to learn less or even not learn from the global model. Then, the client can use the testing accuracy $acc_{ref}$ as a reference to flexibly adjust $\lambda_{ditto}$ by Eq.~\eqref{eq: adaptiveditto}. In the formula, $\lambda_{max}$ and $\lambda_{min}$ are the maximum and minimum values for $\lambda_{ditto}$, $\eta_{ditto}$ is the learning rate, $acc_{local}$ is the testing accuracy of the local model, and $acc_{thres}$ is the minimum threshold to increase $\lambda_{ditto}$. $\lambda_{max}$ and $\lambda_{min}$ restrain the coefficient of the $l2$-regularization in a reasonable interval. $\eta_{ditto}$ controls each mini-batch iteration's growing/decaying speed for $\lambda_{ditto}$. $acc_{thres}$ is the threshold to control whether the current global model is worth being learned. In other words, the local model will absorb from the global model, only if $acc_{global}$ is higher than $acc_{local}+acc_{thres}$. The client secretly shares his/her update with servers after completing the training. The algorithm of FLVoogd's client is manifested in Alg.~\ref{alg: flvoogd_client_1}.

\begin{equation}
    \label{eq: adaptiveditto}
    \lambda_{ditto} := \min(\lambda_{max}, \max(\lambda_{min}, \lambda_{ditto}+\eta_{ditto}(acc_{ref}-acc_{local}-acc_{thres})))
\end{equation}

\begin{algorithm}[htb]
    \caption{FLVoogd client algorithm}
	\label{alg: flvoogd_client_1}
	\begin{algorithmic}[1]
	\State \textbf{Input}: 
	\State $\mathcal{D}_{train}$, $\mathcal{D}_{eval}$ \Comment{$\mathcal{D}_{train}$ is the training set, $\mathcal{D}_{eval}$ is the testing set}
	\State $W_{local}$, $W_{global}$, $F$ \Comment{$W_{local}$/$W_{global}$ are the local/global parameters, $F$ is the objective function}
	\State $\eta_{local}$, $\eta_{global}$, $E$ \Comment{$\eta_{local}$/$\eta_{global}$ is the learning rate for the local/global model, $E$ is the number of local training epochs}
	\State $\lambda_{ditto}^{(0)}$, $\eta_{ditto}$, $acc_{thres}$, $\lambda_{min}$, $\lambda_{max}$ \Comment{$\lambda_{ditto}^{(0)}$ is the initial value for $\lambda_{ditto}$, $\eta_{ditto}$ is the learning rate for $\lambda_{ditto}$, $acc_{thres}$ is the threshold to start learning, $\lambda_{min}$/$\lambda_{max}$ is the minimum/maximum learning rate of $\lambda_{ditto}$}
	\State $\tilde{w}_{global}\gets{client(receive)}$ \Comment{receive the update from the server}
	\State $W_{global}\gets{W_{global}+\tilde{w}_{global}}$
	\State $W_{temp}\gets$deepCopy($W_{global}$)
	\State $acc_{ref}\gets$Eval($W_{global}$,$\mathcal{D}_{eval}$)
	\For{local epoch $e$: $1,2,...,E$}
    	\For{batch iteration $\mathcal{B}\in\mathcal{D}_{train}$ }
        \State $W_{global}\gets{W_{global}-\eta_{global}\triangledown{F}(W_{global},\mathcal{B})}$ \Comment{global train}
        \State $W_{local}\gets{W_{local}-\eta_{local}(\triangledown{F}(W_{local},\mathcal{B})+\lambda_{ditto}(W_{local}-W_{global}))}$ \Comment{local train}
        \State $acc_{local}\gets$Eval($W_{local}$,$\mathcal{D}_{eval}$)
        \State $\lambda_{ditto}\gets\lambda_{ditto}+\eta_{ditto}(acc_{ref}-acc_{local}-acc_{thres})$ by Eq.~\eqref{eq: adaptiveditto}
        \EndFor
	\EndFor
	\State $w\gets{W_{global}-W_{temp}}$
	\State $||w||\gets{\sqrt{\sum_jw^2_j}}$, $\overline{w}\gets\frac{w}{||w||}$, $\overline{v}\gets{\frac{\{w_j:j=k,...,m\}}{\sqrt{\sum_{j=k}^mw^2_j}}}$ \Comment{$k$ is the starting index of the last layer}
    \State $client(send)\gets{\overline{w}, ||w||, \overline{v}}$
	\end{algorithmic}  
\end{algorithm}

%% file: sections/experiment.tex
\section{Experiment}
\subsection{Experimental Setup}
We conducted all the experiments using PyTorch, and the source code was available on GitHub\footnote{\href{https://github.com/anonymous/anonymous}{https://github.com/anonymous/anonymous}}.

\textbf{Datasets and Neural Network}. We followed the recent research on Byzantine attacks~\cite{3-poisonattack, attackTails} on FL and chose a typical application scenario - image classification. The datasets in our experiments included MNIST, CIFAR-10, and EMNIST. The non-IID data splitting for MNIST and EMNIST in our experiment followed the method carried out in FLTrust~\cite{FLTrust}, where $Deg_{nIID}$ ($q$ in~\cite{FLTrust}) controlled the level of non-IID. In terms of EMNIST, we applied the method from~\cite{3-fedml} where the smaller $\alpha_{sim}$ was, the more tasks were dissimilar. CNNs were used as our global and local models, where CIFAR-10 was trained by ResNet-20~\cite{2-ResNet} (269,772 parameters in total), and MNIST \& EMNIST were trained by a $2\times$convolutional layers' NN. ResNet-20 was a pre-trained version\footnote{\href{https://github.com/chenyaofo/pytorch-cifar-models}{https://github.com/chenyaofo/pytorch-cifar-models}} to accelerate the training process.

\textbf{Evaluation Metrics}. Main Task Accuracy (MA) represents the accuracy of a model tested by its benign task. It indicates the fraction of correct predictions. If the model is under targeted attacks, Backdoor Accuracy (BA) is the metric to reflect how successful the adversaries are. It denotes the fraction of correct predictions for backdoor samples.

\textbf{FL Configuration}. The total number of clients $N$ was set to 100. Each client received unique training samples and testing samples from the split.  The learning rates of global model $\eta_{global}$ and local model $\eta_{local}$ were 0.01. The local training epoch $E$ was 1 since clients did not hold an adequate number of samples. The coefficient of $l2$-regularization $\lambda_{ditto}$ was initialized as 0 and its min-max interval was $[0.0,2.0]$, where the maximum was suggested by~\cite{ditto}. The threshold $acc_{thres}$ for the local model starting learning from the global was 0.05. The learning rate $\eta_{ditto}$ for $\lambda_{ditto}$ was 1. The expected clipping ratio $\gamma$, the initial clipping boundary $c^{(0)}$, the clipping learning rate $\eta_{c}$ were 0.5, 10, 0.3. Other settings varied for different experiments.

\textbf{Malicious Configuration}. Both model poisoning attacks and data poisoning attacks share a parameter Poisoned Model Rate (PMR), indicating the fraction of poisoned models \textbf{for each round}. If the attacking type is data poisoning, it has one more parameter Poisoned Data Rate (PDR), representing the poisoned ratio of the data. Common attacking settings are shown in table~\ref{tab: attacking_setting}, where settings are nearly marginal thresholds, below which the attacking effect is non-significant even if it escapes from being detected.

\begin{table}[htb]
\small
     \caption{Attack settings}
     \label{tab: attacking_setting}
\begin{tabular}{|c|cc|}
\hline
Attacks & \multicolumn{1}{c|}{(E)MNIST}                                                                                                                & CIFAR-10                                                                                                                     \\ \hline
A1      & \multicolumn{2}{c|}{mean = 0, std = 1}                                                                                                                                                                                                                                      \\ \hline
A2      & \multicolumn{2}{c|}{Krum's $\epsilon$ = $10^{-3}$, threshold = $2\times10^{-2}$}                                                                                                                                                                                          \\ \hline
A5      & \multicolumn{1}{c|}{\begin{tabular}[c]{@{}c@{}}a 5x5 white square is inserted into \\ the targeted data and labeling it as ``0''\end{tabular}} & \begin{tabular}[c]{@{}c@{}}the semantic pattern is a car \\ with stripes and labeling ``car'' to ``bird''\end{tabular}           \\ \hline
A6      & \multicolumn{1}{c|}{\begin{tabular}[c]{@{}c@{}}by adding Ardis\_IV to \\ training and labeling ``7'' as ``1''\end{tabular}}                      & \begin{tabular}[c]{@{}c@{}}by adding Southwest Airline images to \\ training and labeling ``airplane'' as ``truck''\end{tabular} \\ \hline
\end{tabular}
\end{table}

\subsection{Fending-off Byzantine Attacks}
\label{subsec: byzantine_defense}
\begin{figure}[htb]
    \centering
    \subfigure[\scriptsize{A1: PMR=19/40}\label{fig: 5-mnist-A1}]{\includegraphics[width=0.3\textwidth]{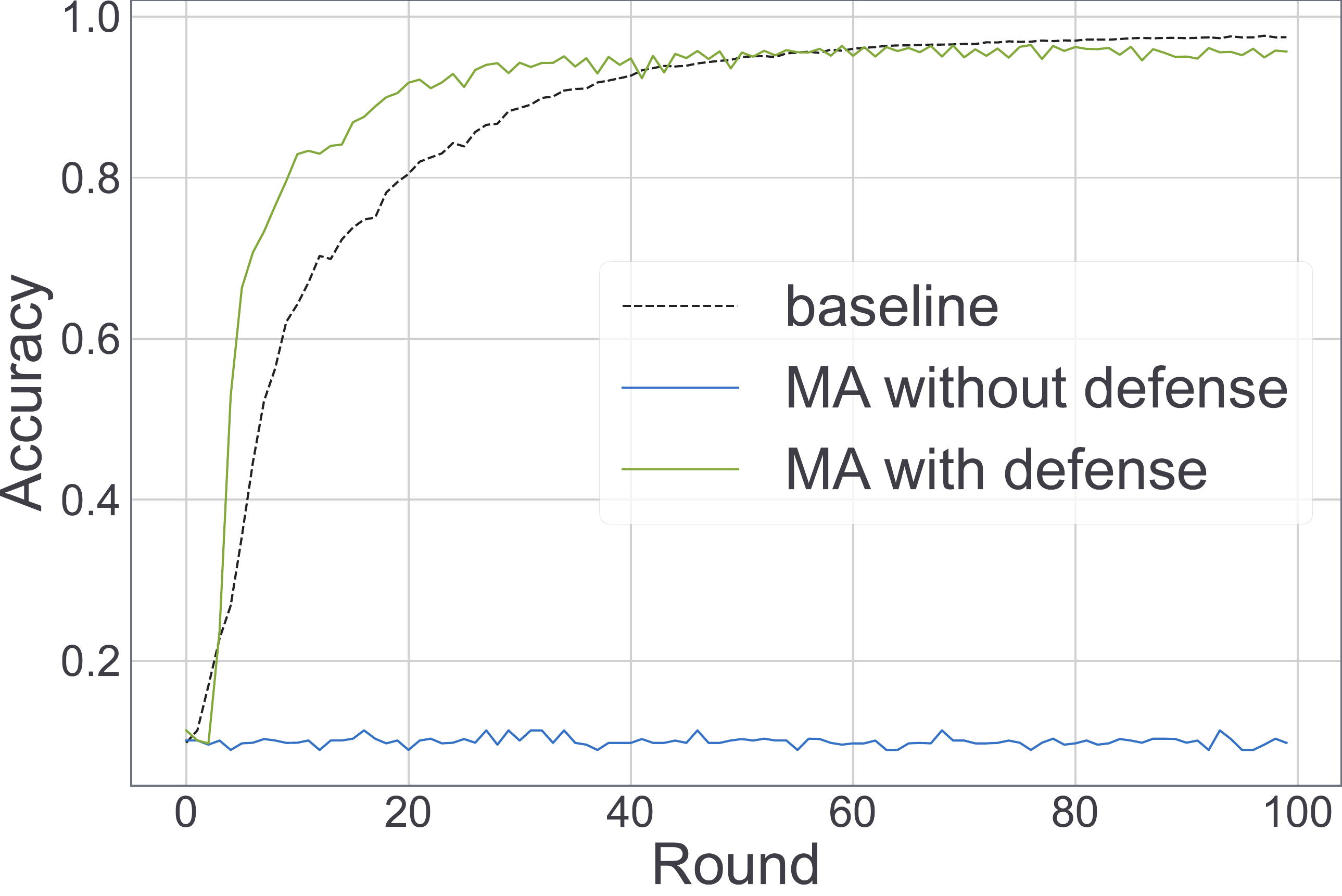}}
    \hfill
    \subfigure[\scriptsize{A2: PMR=19/40}\label{fig: 5-mnist-A2}]{\includegraphics[width=0.3\textwidth]{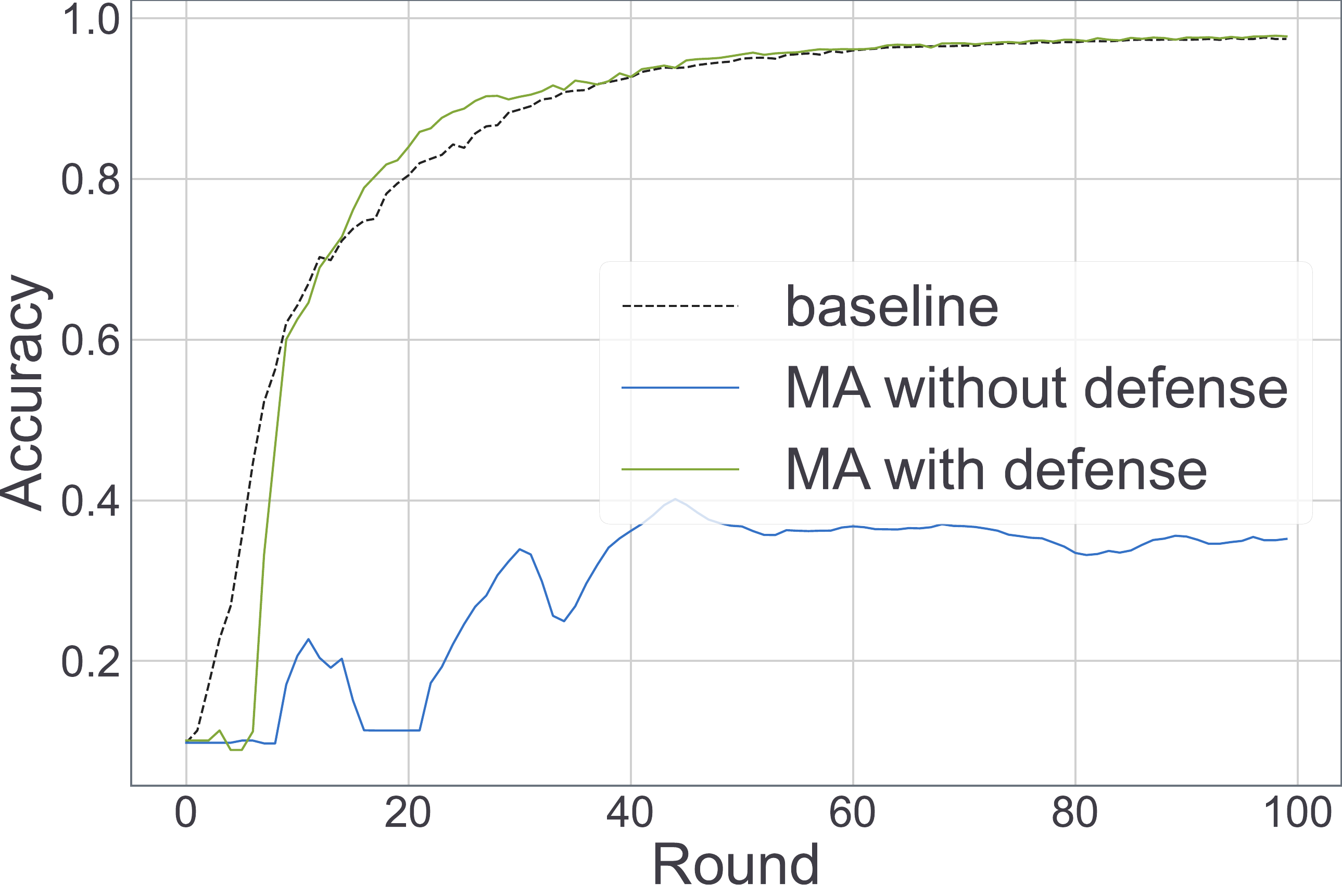}}
    \hfill
    \subfigure[\scriptsize{A3: PMR=19/40}\label{fig: 5-mnist-A3}]{\includegraphics[width=0.3\textwidth]{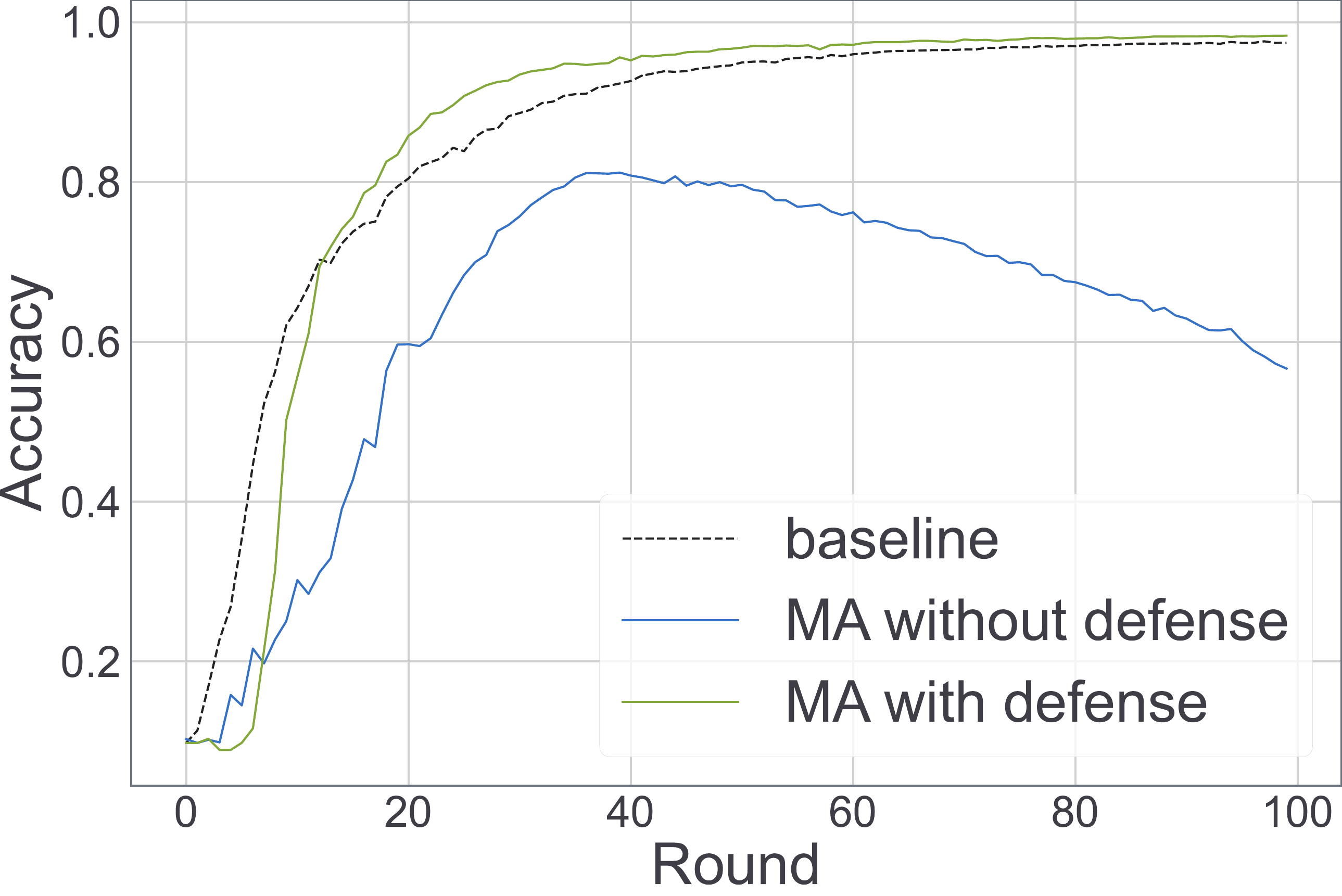}}
    \hfill
    \subfigure[\scriptsize{A4: PMR=19/40, PDR=1.0}\label{fig: 5-mnist-A4}]{\includegraphics[width=0.3\textwidth]{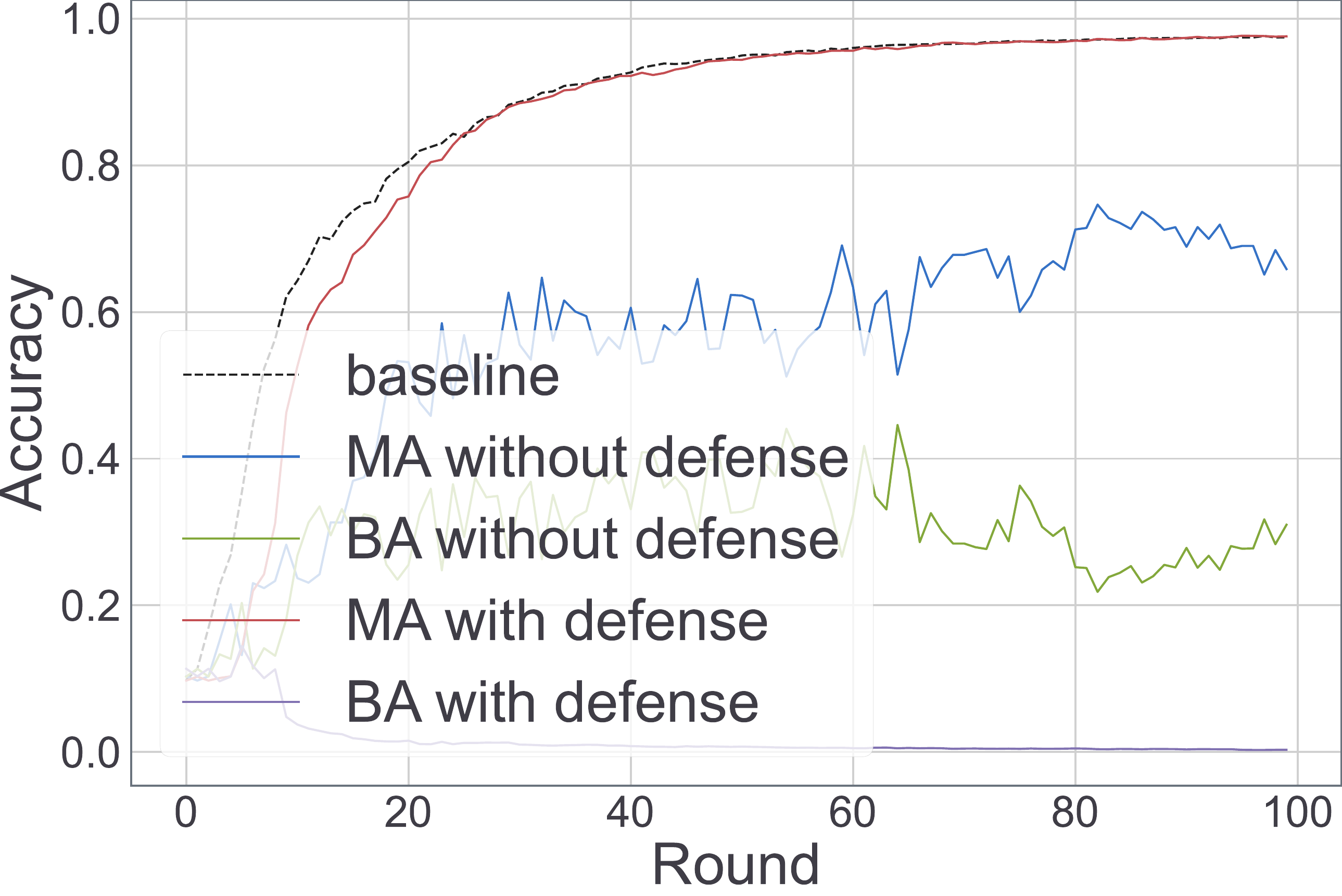}}
    \hfill
    \subfigure[\scriptsize{A5: PMR=19/40, PDR=0.2}\label{fig: 5-mnist-A5}]{\includegraphics[width=0.3\textwidth]{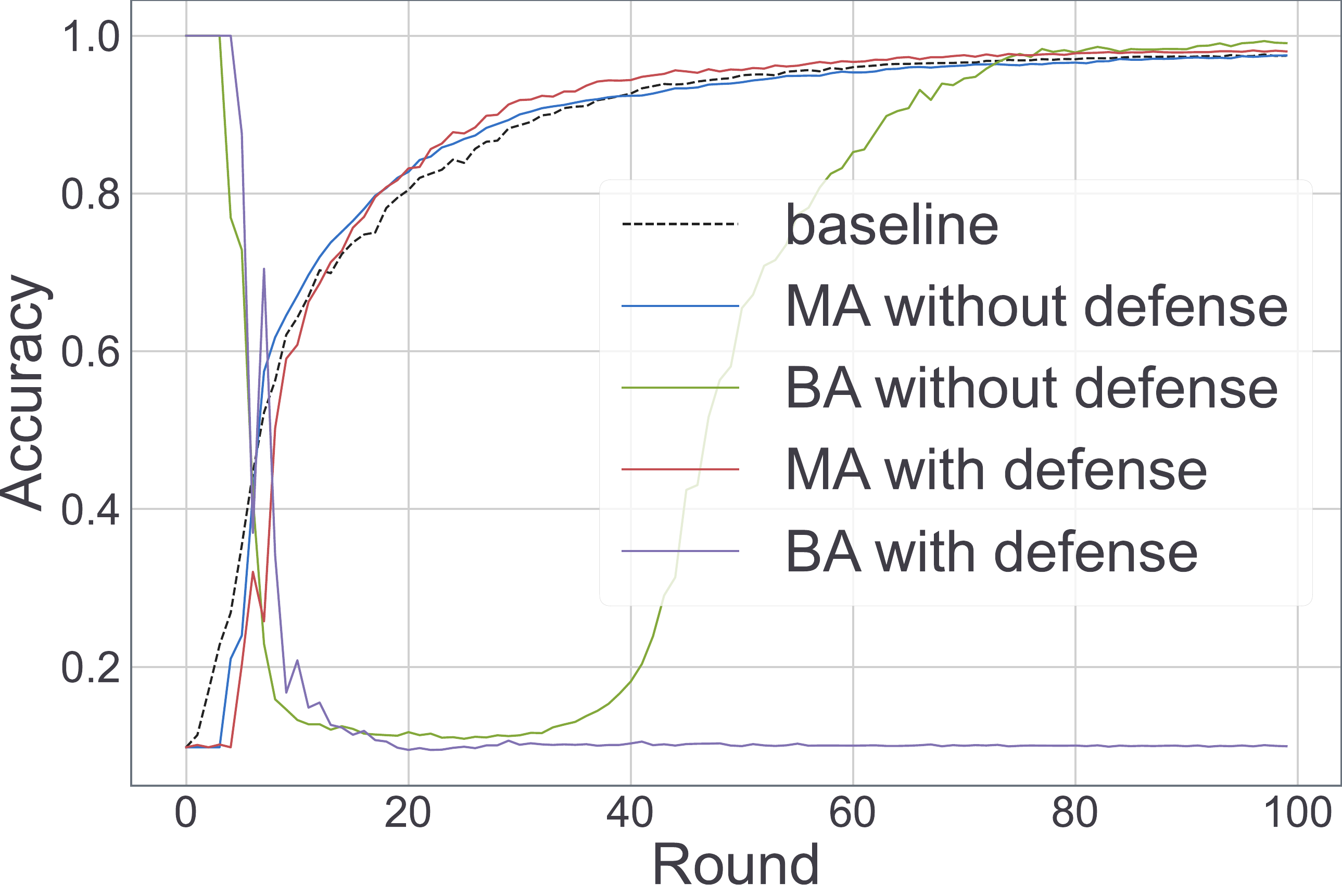}}
    \hfill
    \subfigure[\scriptsize{A6: PMR=19/40, PDR=0.33}\label{fig: 5-mnist-A6}]{\includegraphics[width=0.3\textwidth]{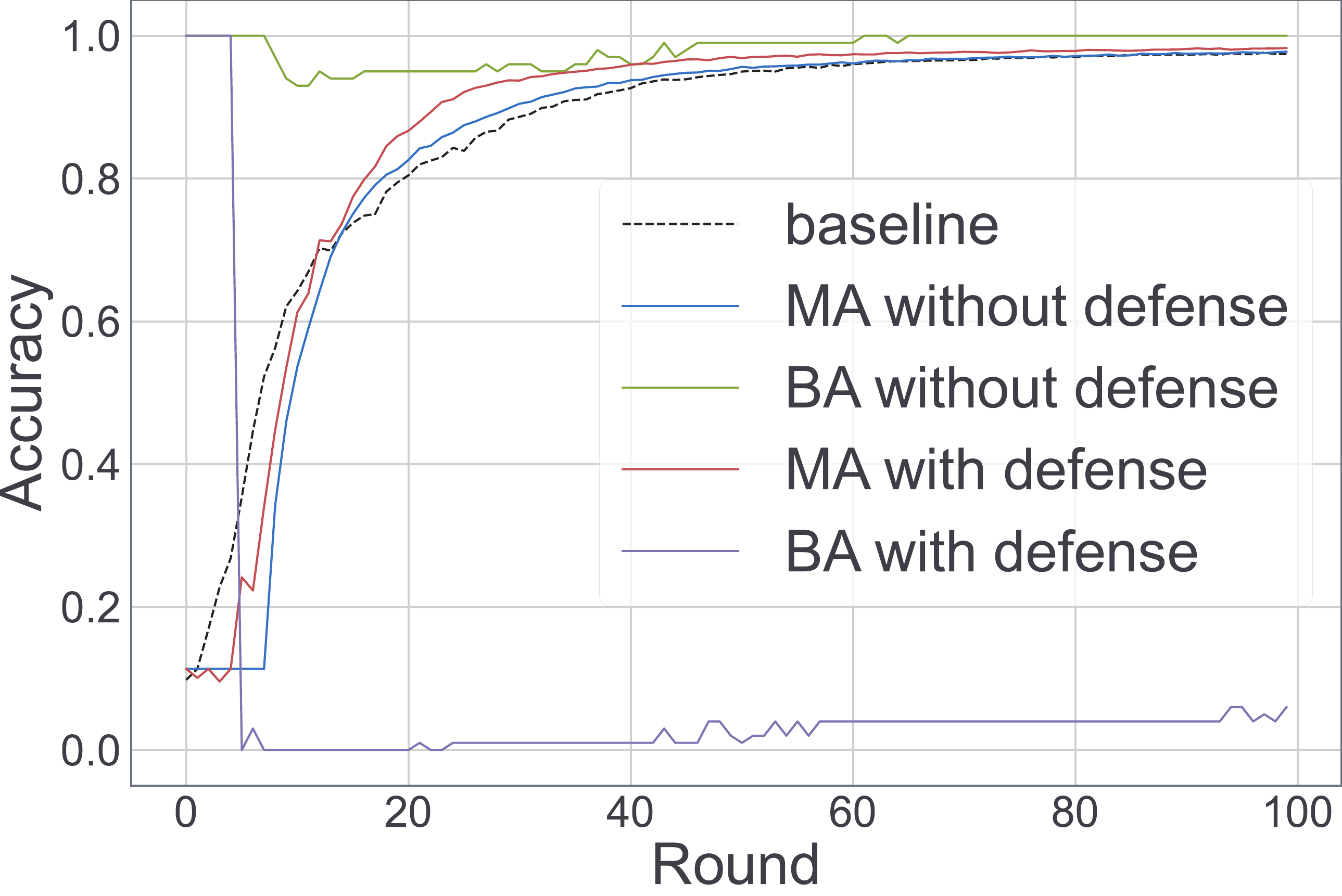}}
    \caption{Byzantine attacks on MNIST}
    \label{fig: 5-mnist-byzantine}
\end{figure}

Fig.~\ref{fig: 5-mnist-byzantine} shows MNIST under different attacks from A1 to A6. ``Baseline'' or ``without defense'' results from the server ruining naive aggregation. 21 clients are uniformly randomly selected from 100 participants for the baseline, while 40 clients for other situations. In Fig.~\ref{fig: 5-mnist-A1}, due to malicious random uploads, the aggregated updates are meaningless, leading to the blue curve with extremely low accuracy; however, the filtering process conducts so effectively that the learning curve - the green one - can behave normally under this attack. In Fig.~\ref{fig: 5-mnist-A2}, the Krum attack tries its best to upload counter-directional updates to devalue the global accuracy. Consequently, the global accuracy is even worse than a random guess (50\%). However, the global accuracy can reach an original level using the defense of FLVoogd. In Fig.~\ref{fig: 5-mnist-A3}, the trimmed-mean attack starts to degrade the model accuracy approximately at midterm and reduces accuracy from higher than 80\% to less than 60\% within 50 rounds. FLVoogd prevents this malicious reduction of accuracy effectively. In Fig.~\ref{fig: 5-mnist-A4}, Byzantine clients flip the labels of all training samples, rendering the final prediction like a random guess. One notable point is that the filter cannot correctly recognize flipping uploads in several initial rounds, but it can acknowledge and expel malicious updates once the global model learns a little from those majorities. In Fig.~\ref{fig: 5-mnist-A5}, the backdoor triggering attack shows its supremacy at the initial stage, and BA can easily approximate to 100\% in the first several rounds. After the model learns sufficient benign samples, BA tends to decrease while MA tends to increase. Without the defense, BA again grows sharply at midterm and finally attains relatively high accuracy. In contrast, under the protection, BA is restrained at low accuracy and can impossibly revive after the filter starts to work. The filter does not work at the initial stage because the model is chaotic, and the updates produced from the model reveal little information about the data distribution. In Fig.~\ref{fig: 5-mnist-A6}, under no defense, BA manifests likewise A5 at the initial stage but does not decline after MA rises. Since the malicious data is sampled from another dataset without intersection with MNIST, learning from the benign samples cannot benefit the model, so BA persists at high accuracy. However, FLVoogd can successfully detect and block these uploads according to the abnormal data distribution, preventing the intrusion of edge-case.

\begin{figure}[htb]
    \centering
    \subfigure[\scriptsize{A1: PMR=19/40}\label{fig: 5-cifar-A1}]
    {\includegraphics[width=0.3\textwidth]{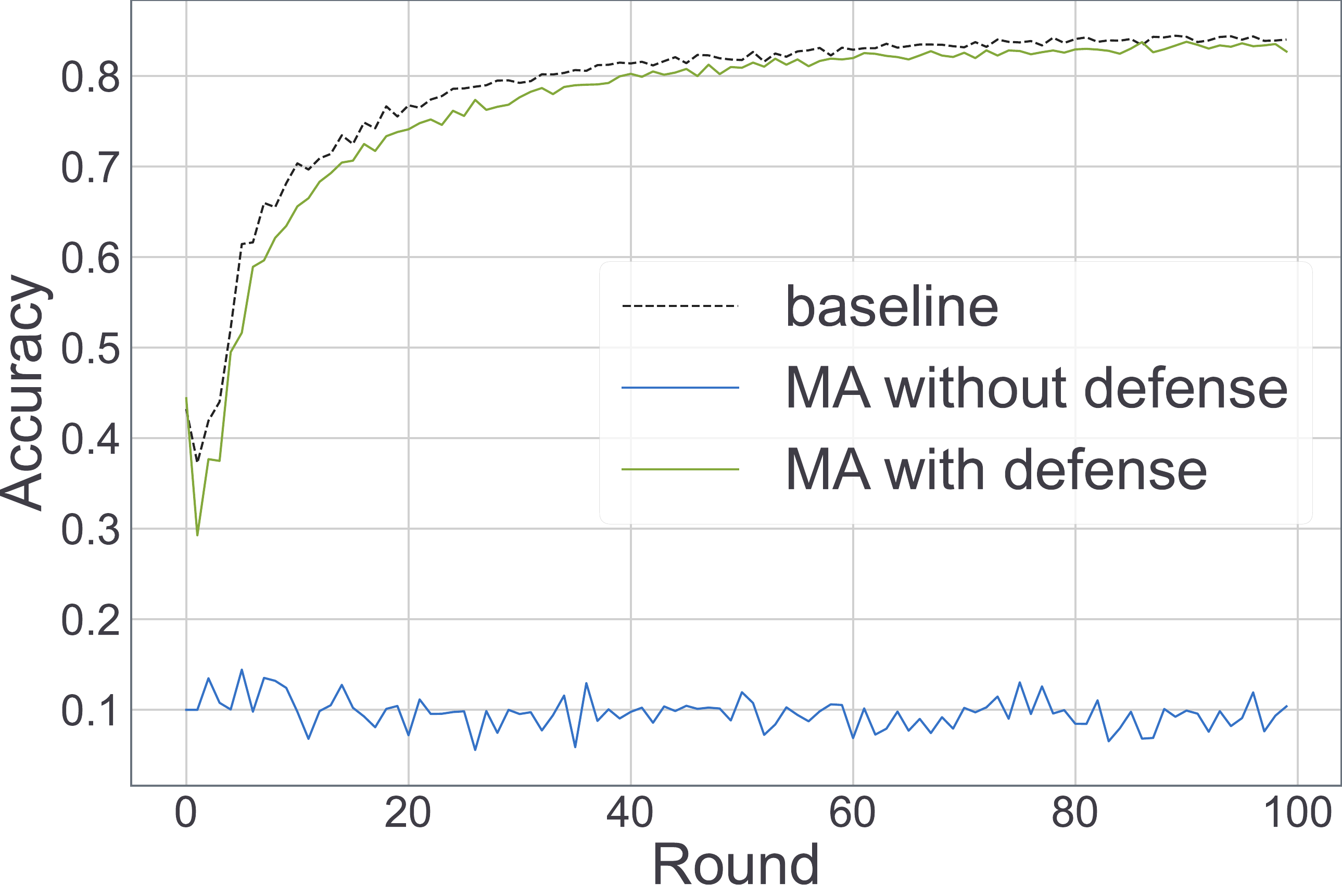}}
    \hfill
    \subfigure[\scriptsize{A2: PMR=19/40}\label{fig: 5-cifar-A2}]
    {\includegraphics[width=0.3\textwidth]{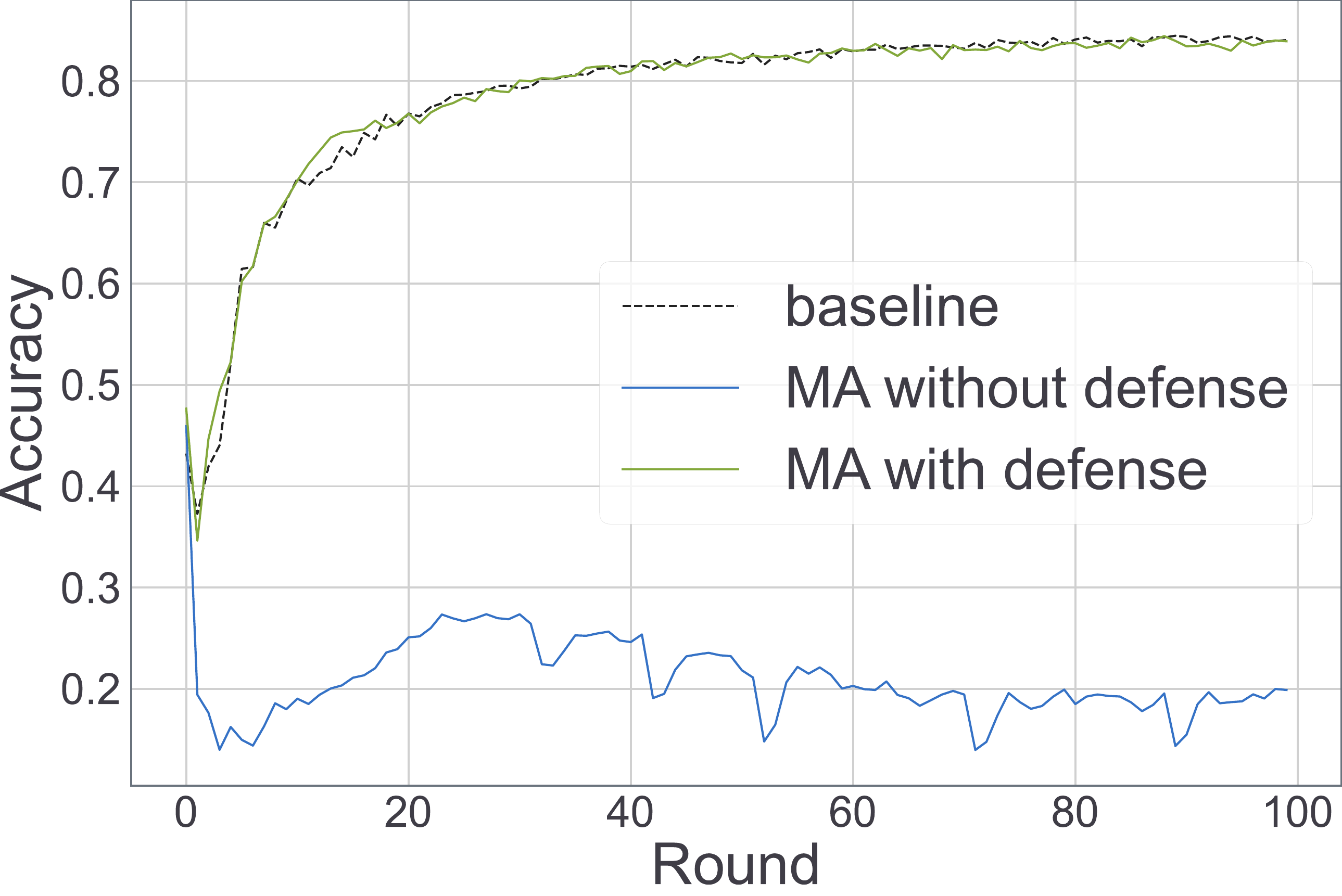}}
    \hfill
    \subfigure[\scriptsize{A3: PMR=19/40}\label{fig: 5-cifar-A3}]{\includegraphics[width=0.3\textwidth]{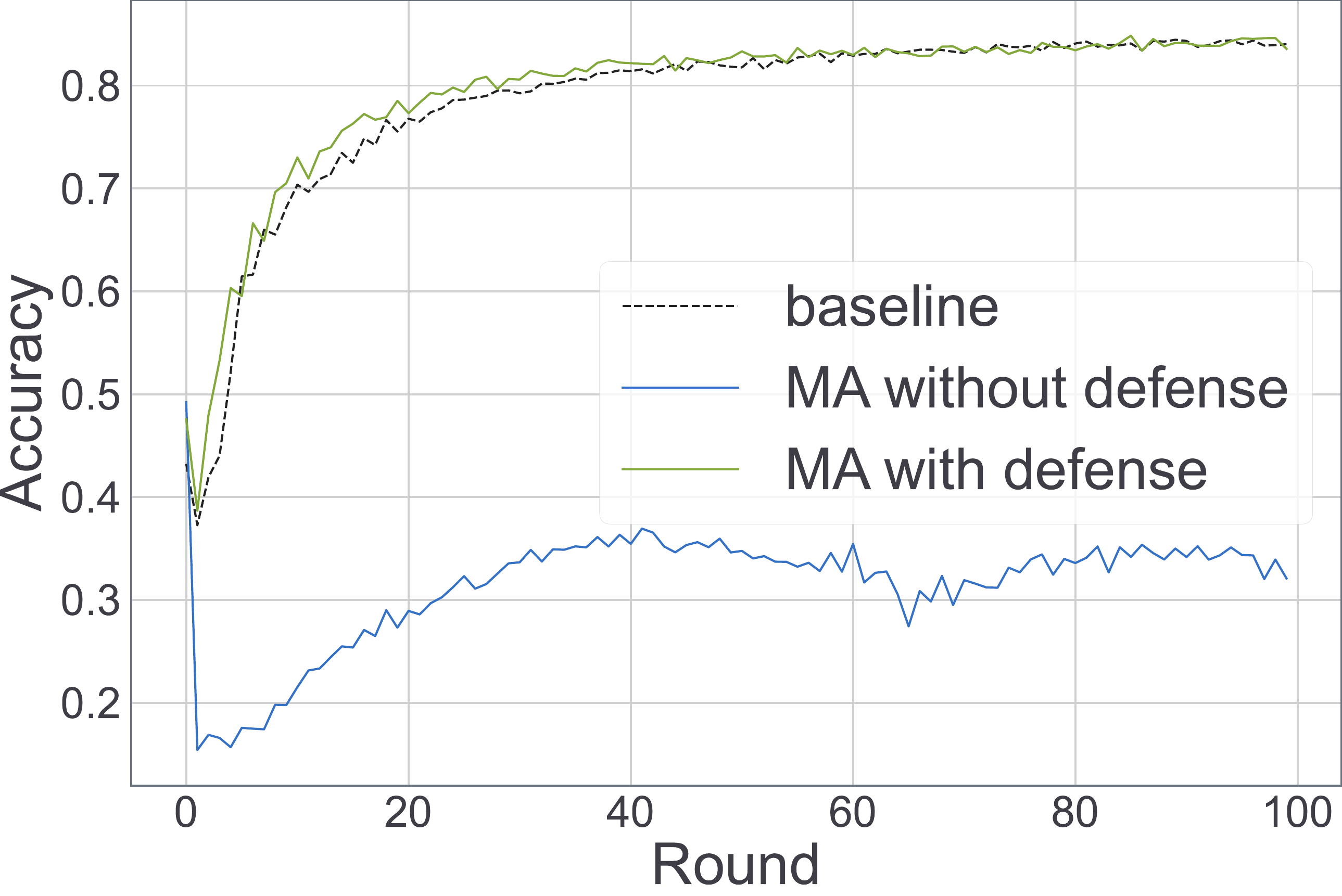}}
    \hfill
    \subfigure[\scriptsize{A4: PMR=19/40, PDR=1.0}\label{fig: 5-cifar-A4}]{\includegraphics[width=0.3\textwidth]{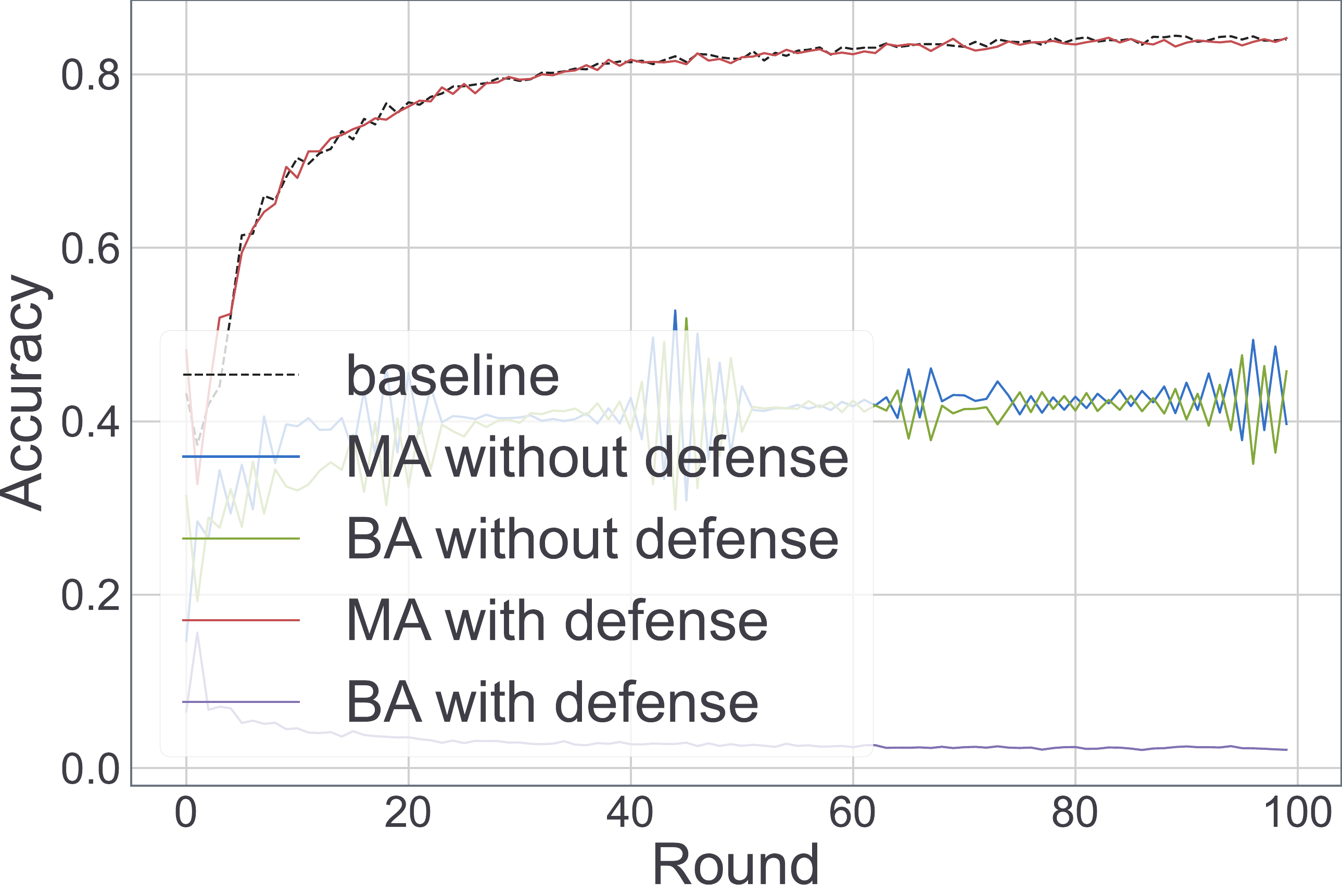}}
    \hfill
    \subfigure[\scriptsize{A5: PMR=19/40, PDR=0.33}\label{fig: 5-cifar-A5}]{\includegraphics[width=0.3\textwidth]{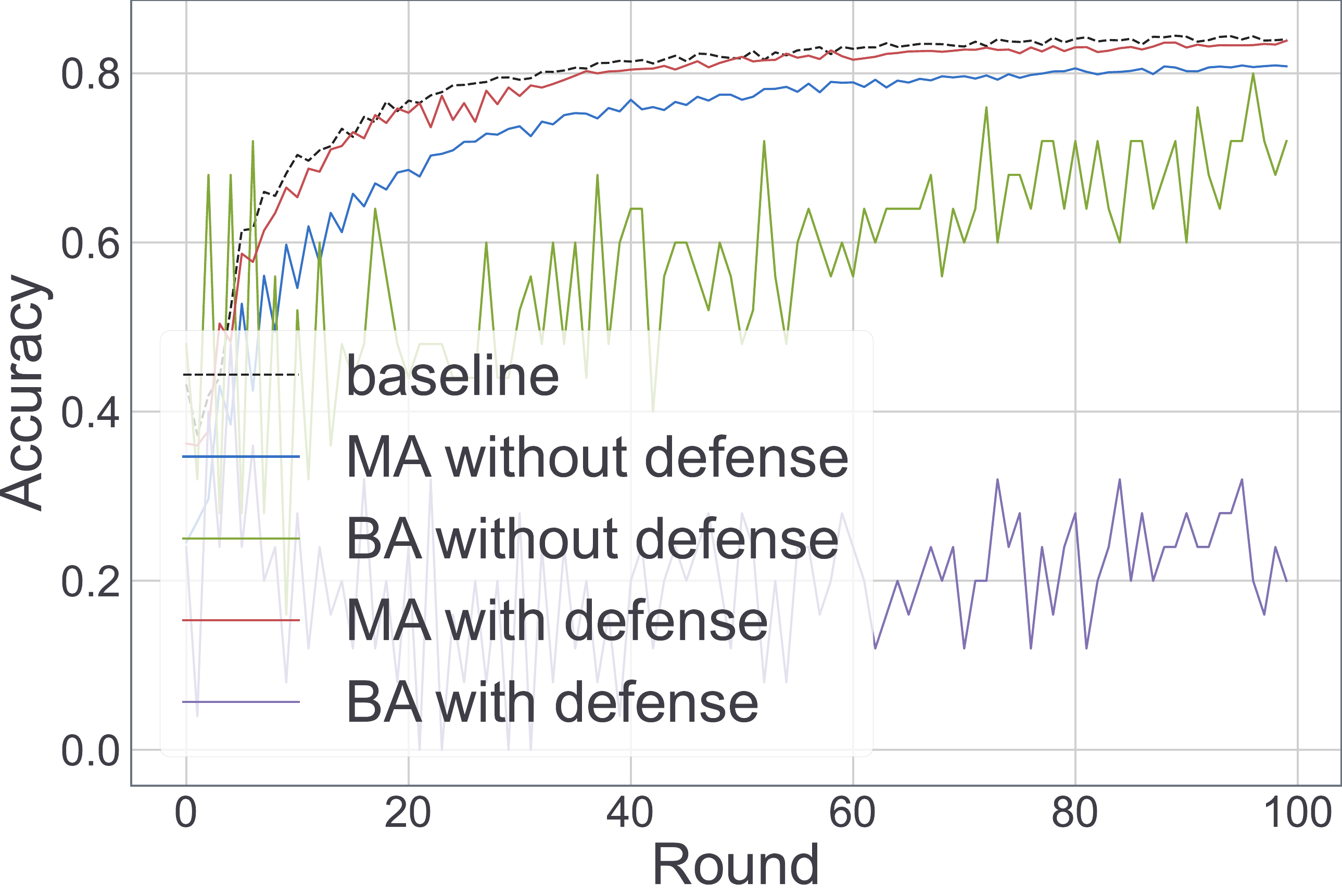}}
    \hfill
    \subfigure[\scriptsize{A6: PMR=19/40, PDR=0.33}\label{fig: 5-cifar-A6}]{\includegraphics[width=0.3\textwidth]{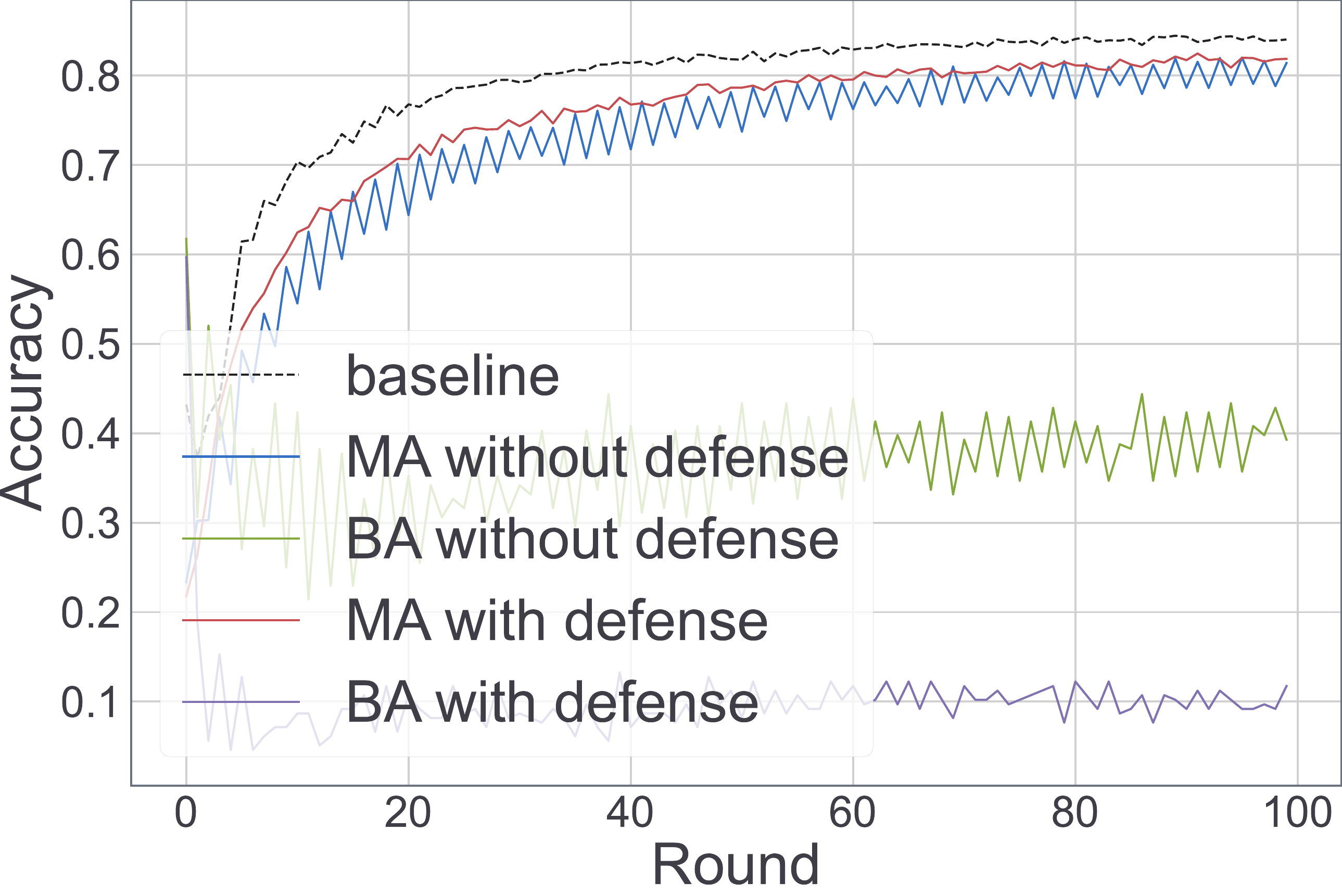}}
    \caption{Byzantine attacks on CIFAR-10}
    \label{fig: 5-cifar-byzantine}
\end{figure}

\begin{figure}[htb]
    \centering
    \subfigure[\scriptsize{A1: PMR=19/40}\label{fig: 5-emnist-A1}]{\includegraphics[width=0.3\textwidth]{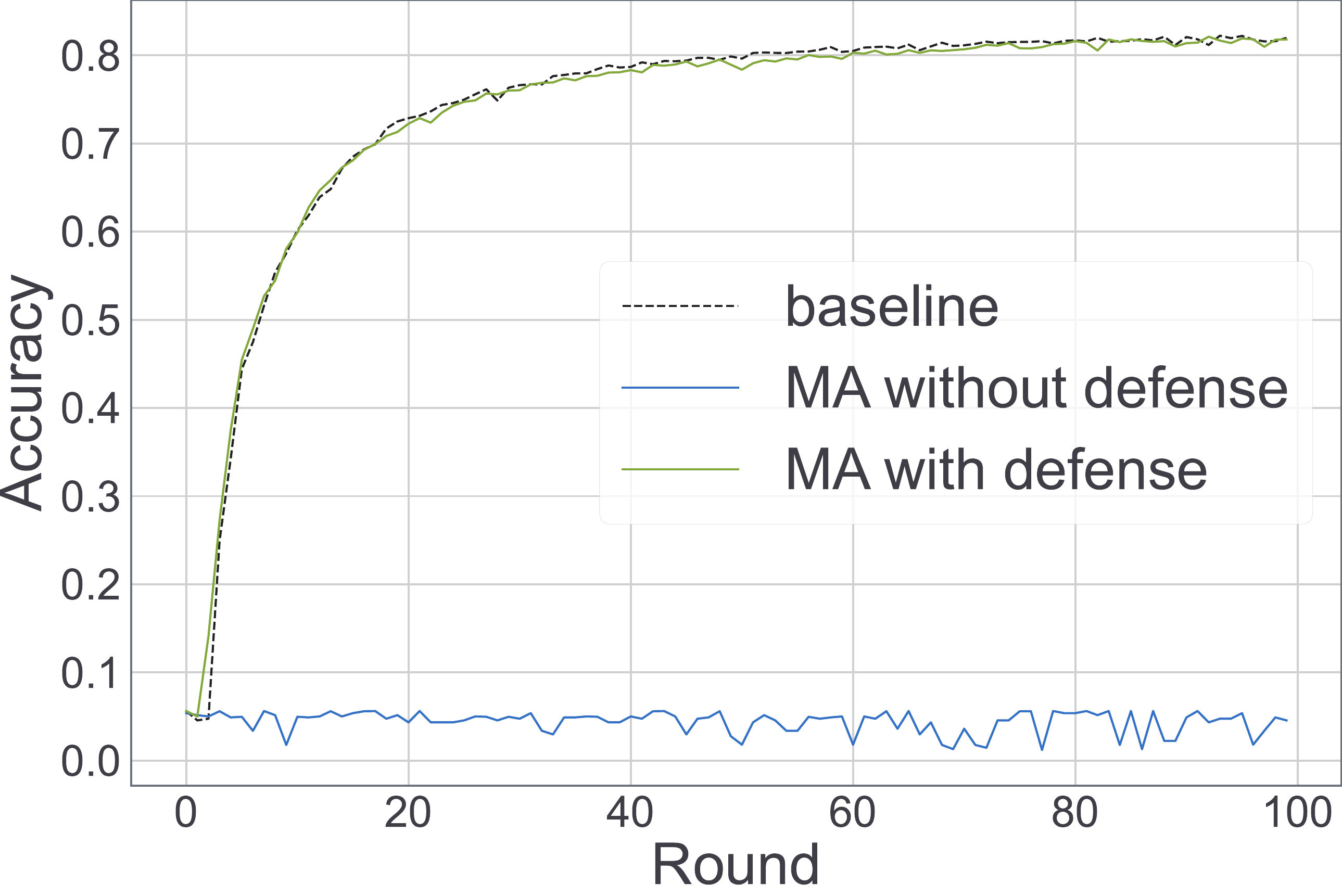}}
    \hfill
    \subfigure[\scriptsize{A2: PMR=19/40}\label{fig: 5-emnist-A2}]{\includegraphics[width=0.3\textwidth]{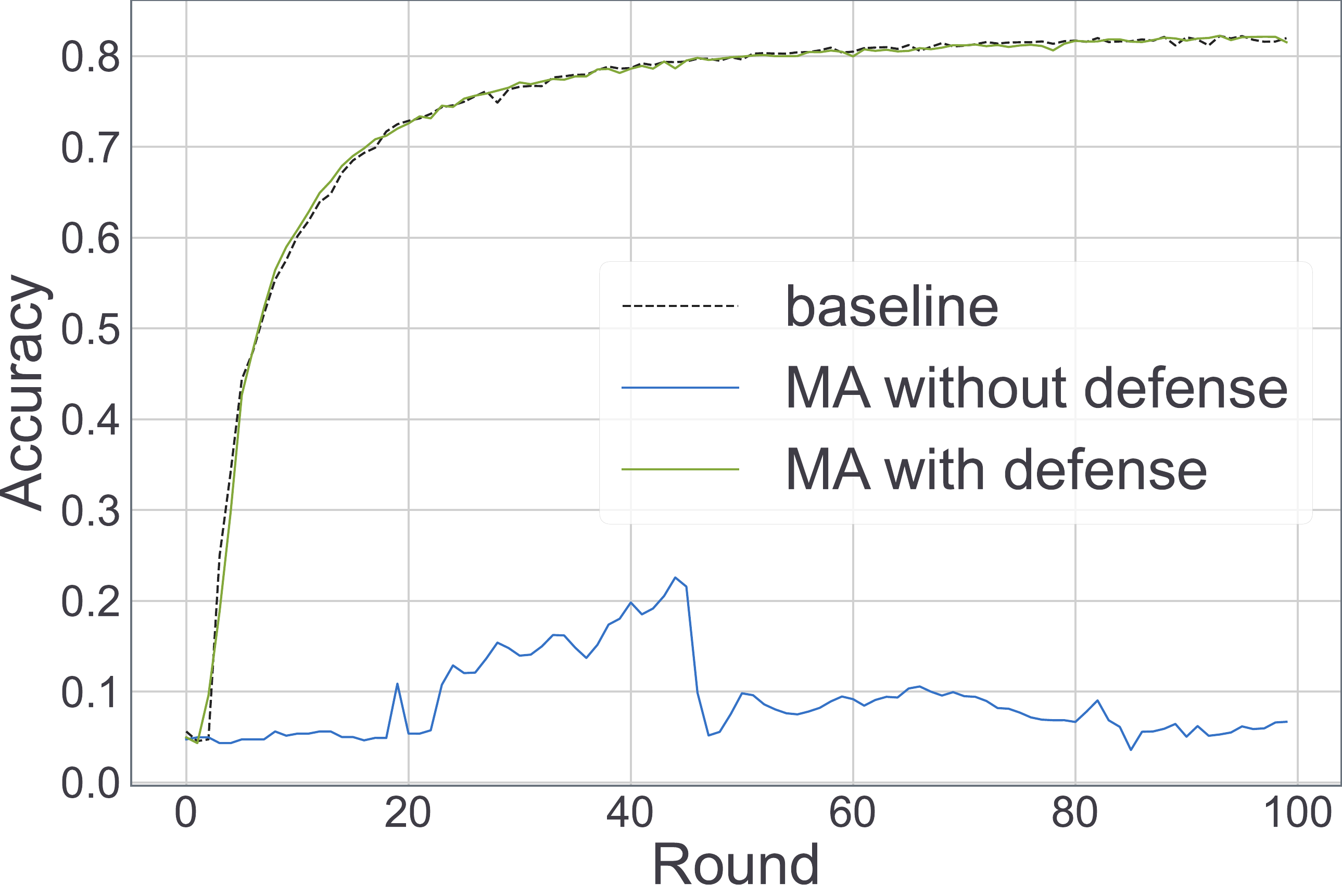}}
    \hfill
    \subfigure[\scriptsize{A3: PMR=19/40}\label{fig: 5-emnist-A3}]{\includegraphics[width=0.3\textwidth]{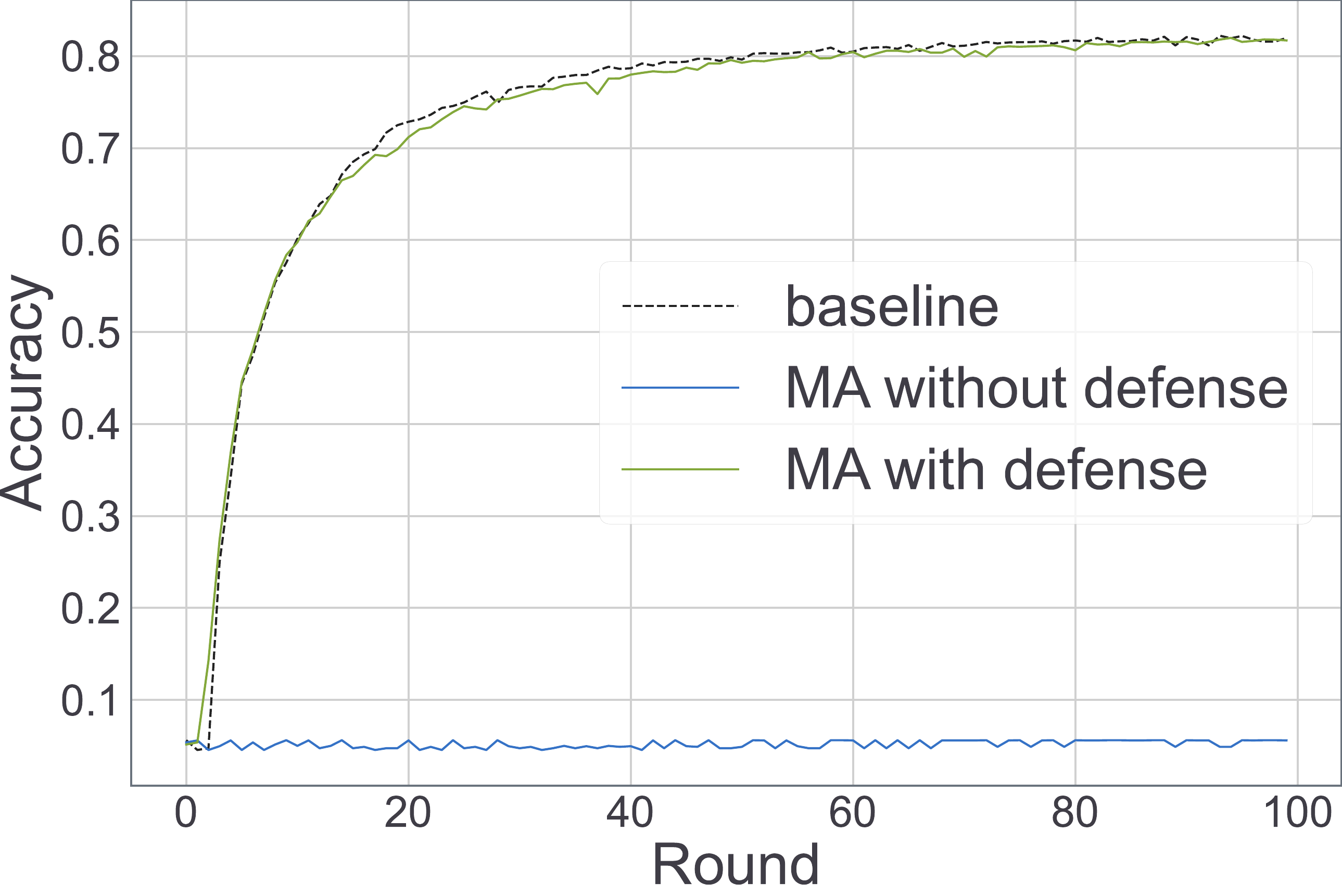}}
    \hfill
    \subfigure[\scriptsize{A4: PMR=19/40, PDR=1.0}\label{fig: 5-emnist-A4}]{\includegraphics[width=0.3\textwidth]{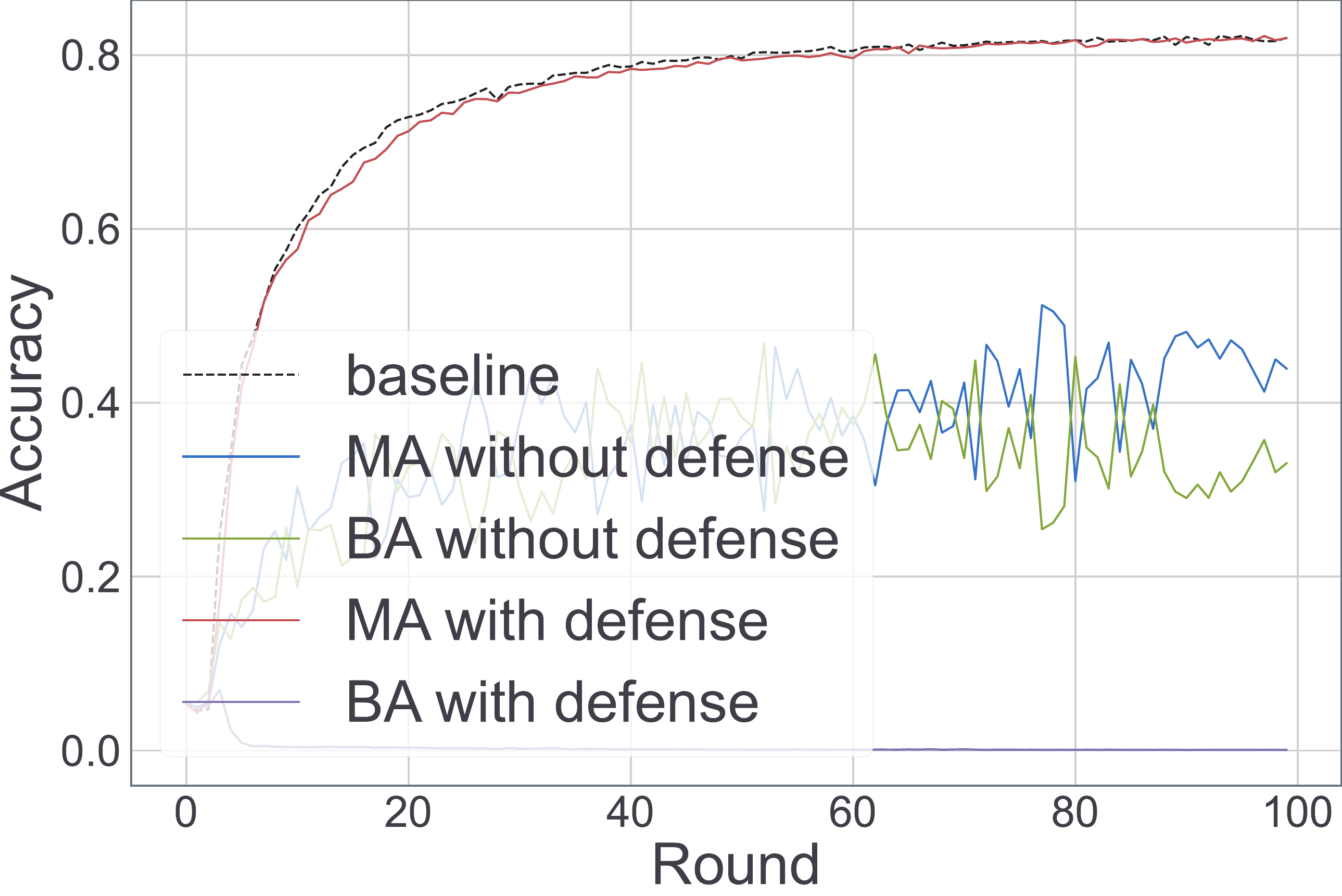}}
    \hfill
    \subfigure[\scriptsize{A5: PMR=19/40}\label{fig: 5-emnist-A5}]{\includegraphics[width=0.3\textwidth]{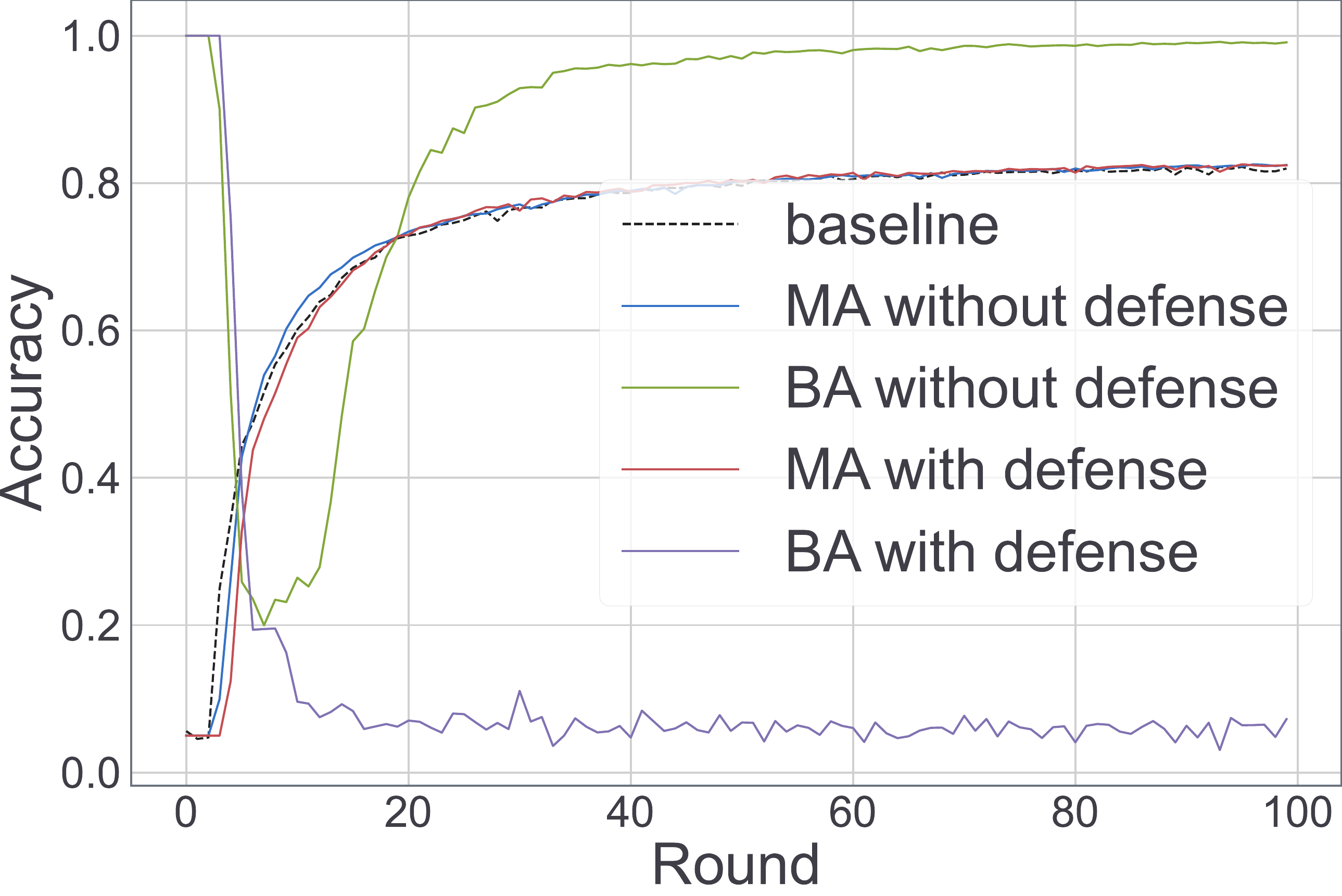}}
    \hfill
    \subfigure[\scriptsize{A6: PMR=19/40, PDR=0.33}\label{fig: 5-emnist-A6}]{\includegraphics[width=0.3\textwidth]{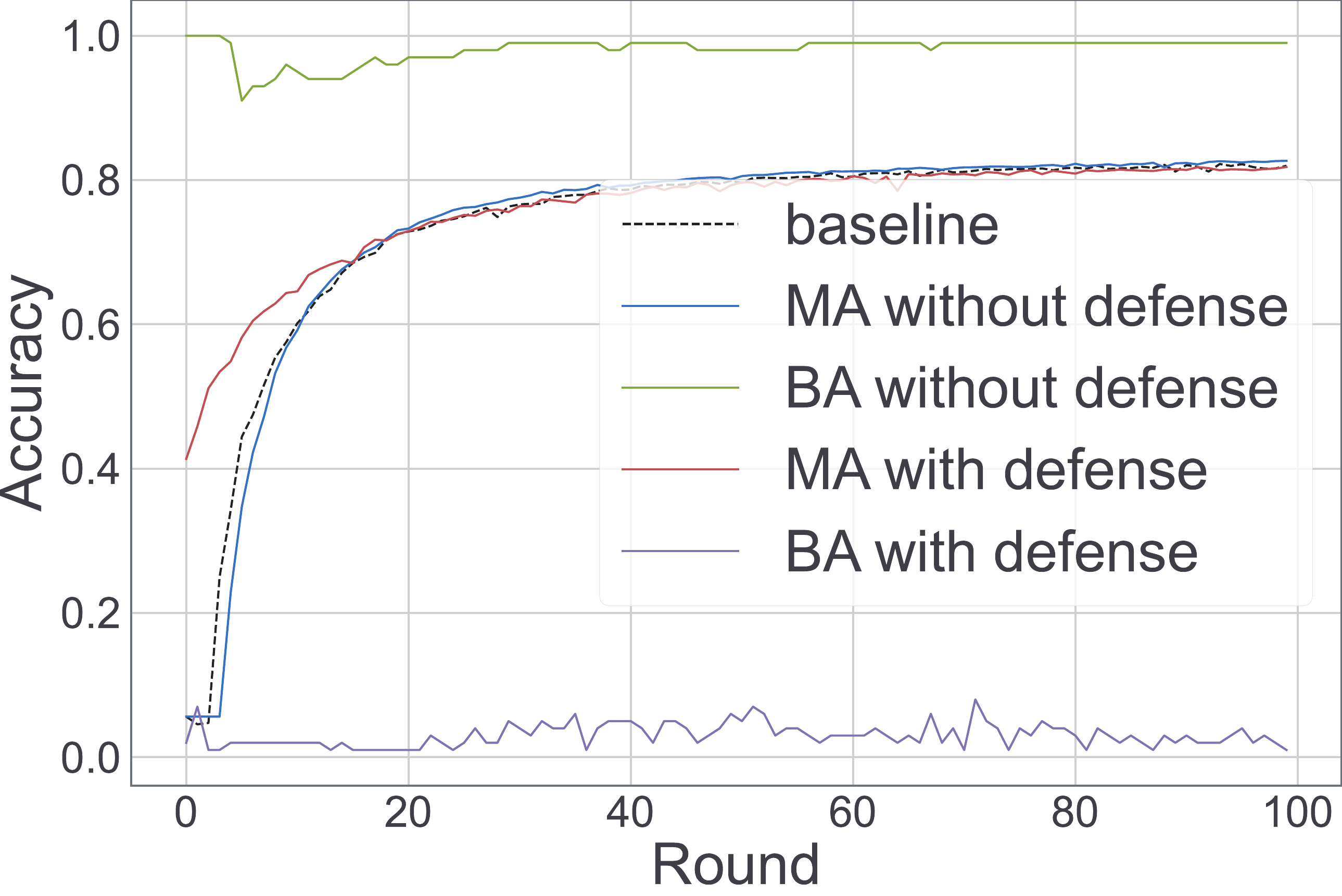}}
    \caption{Byzantine attacks on EMNIST}
    \label{fig: 5-emnist-byzantine}
\end{figure}

Fig.~\ref{fig: 5-cifar-byzantine} shows CIFAR-10 under different attacks from A1 to A6, where all the attacks have been eliminated or restrained below comparably low accuracy. Fig.~\ref{fig: 5-emnist-byzantine} shows EMNIST under various attacks from A1 to A6. Similar to MNIST and CIFAR-10, all the attacks have been eliminated or restrained below comparably low accuracy. However, we initially found that FLVoogd could not prevent EMNIST from A6 productively since 1) we did not use the whole dataset for the training but followed the advice from~\cite{3-fedml}, where 20\% was suggested for 100 clients; 2) there were 62 classes to be classified. Consequently, the model initially required several rounds to figure out what correct ``7'' and ``1'' roughly looked like. The model would not rebound those edge cases if edge-case clients instructed the model incorrectly with the mislabelled pictures at the beginning. Therefore, in the first five rounds, we put the model under training with benign-only samples to compensate for this unfairness. After that, the model could successfully filter out those malicious uploads. 

\subsection{Adding and Tracking DP}
Experiments that test the DP effect and monitor the $\epsilon$ budget will be independently studied in this subsection. The experiments consist of different combinations of subsampling ratio $q$ and noise strength coefficient $\sigma$. The DP noise, to some extent, will adversely affect the convergence and accuracy of the model. In return, this kind of sacrifice gains a differential privacy guarantee. The growth of $\epsilon$ after each iteration is estimated by Moments accountant or R{\.e}nyi-DP (RDP)~\cite{subsampledRDP}, where $\delta$ is set as a constant ($=10^{-3}$~\cite{DPFLclientlevel}) considering 100 is the total number of clients.

\begin{figure}[htb]
    \centering
    \subfigure[\scriptsize{$\sigma=1$, $q=20/100$}\label{fig: 5-mnist-S1}]{\includegraphics[width=0.3\textwidth]{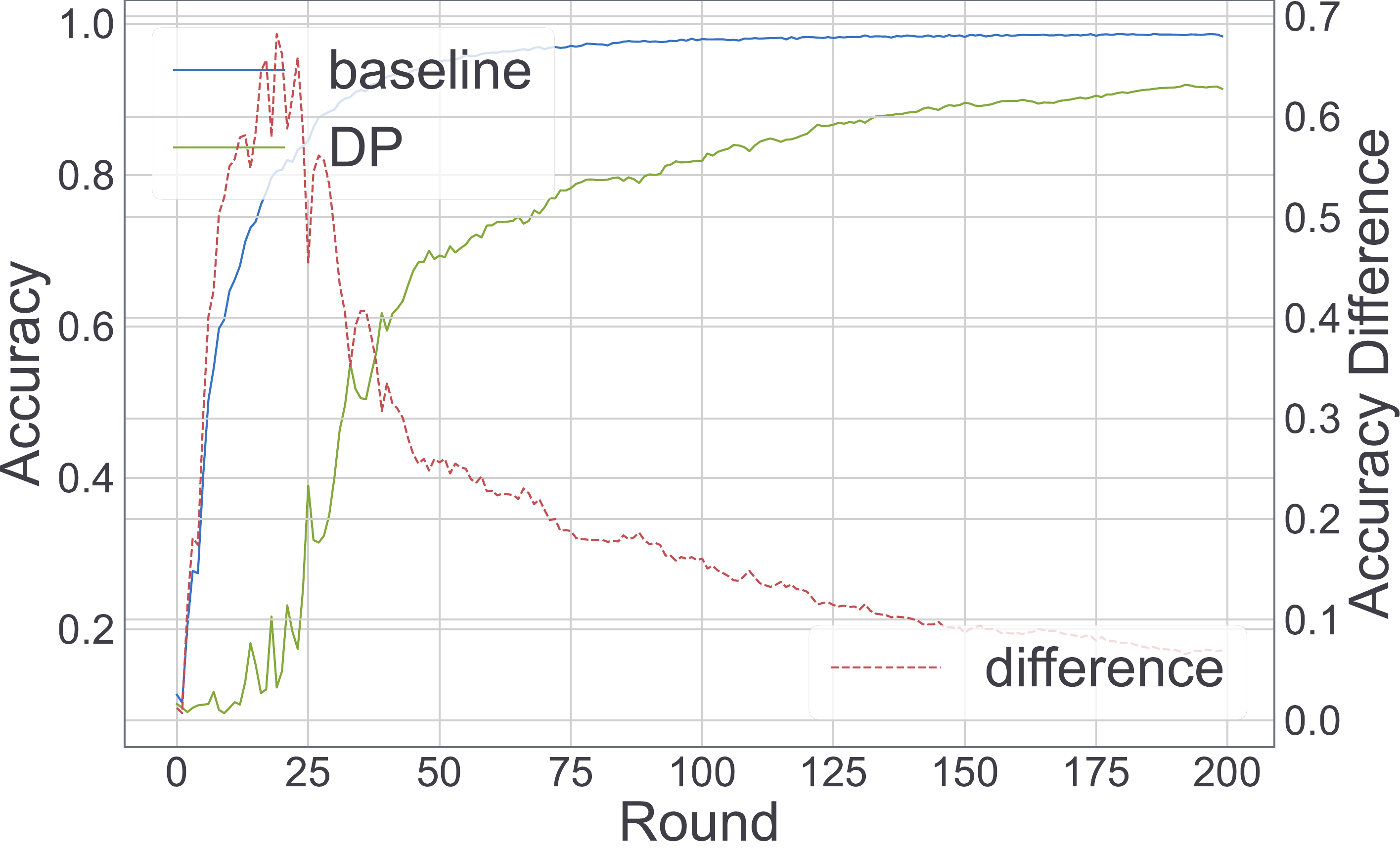}}
    \hfill
    \subfigure[\scriptsize{$\sigma=2$, $q=40/100$}\label{fig: 5-mnist-S2}]{\includegraphics[width=0.3\textwidth]{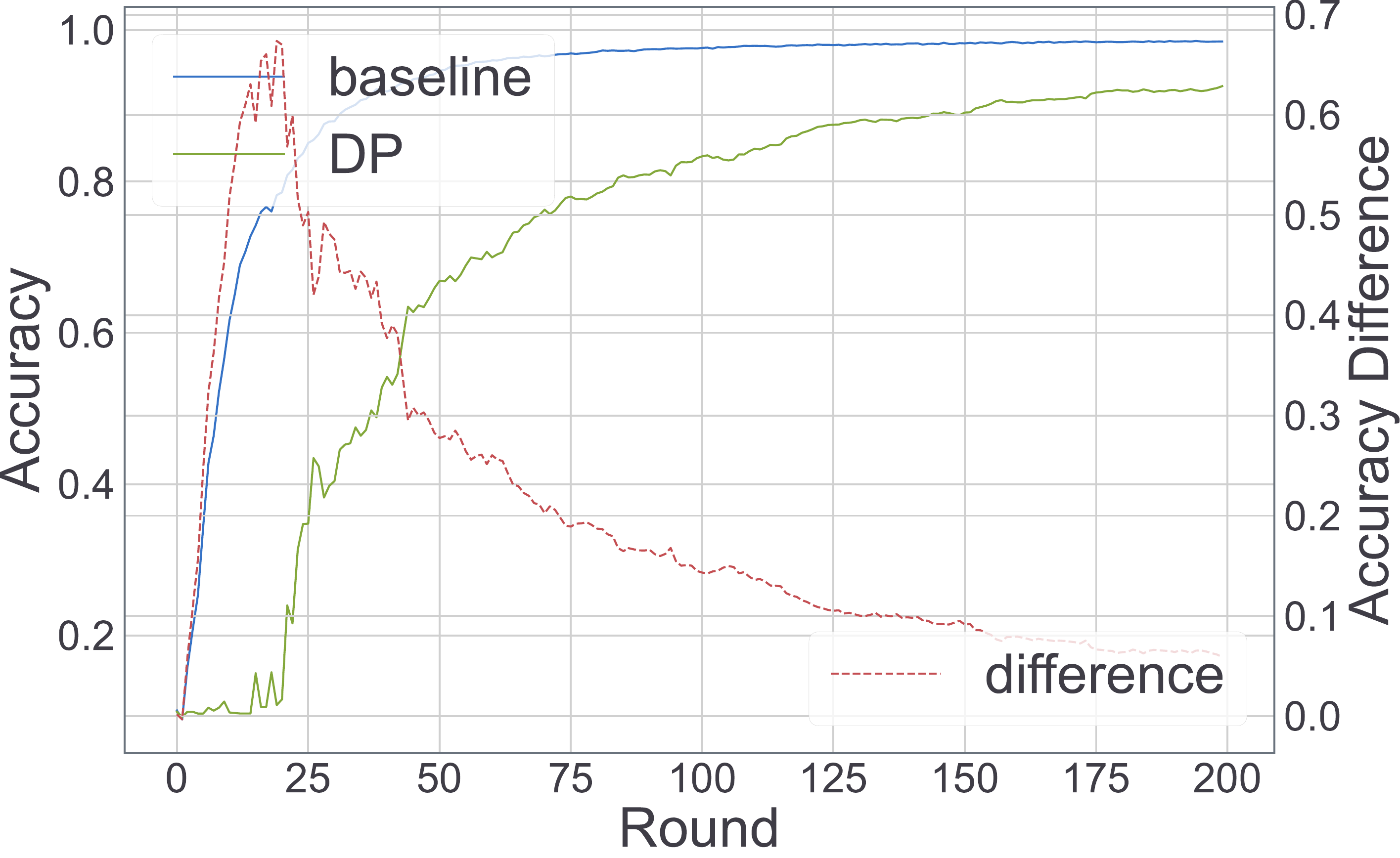}}
    \hfill
    \subfigure[\scriptsize{$\sigma=3$, $q=60/100$}\label{fig: 5-mnist-S3}]{\includegraphics[width=0.3\textwidth]{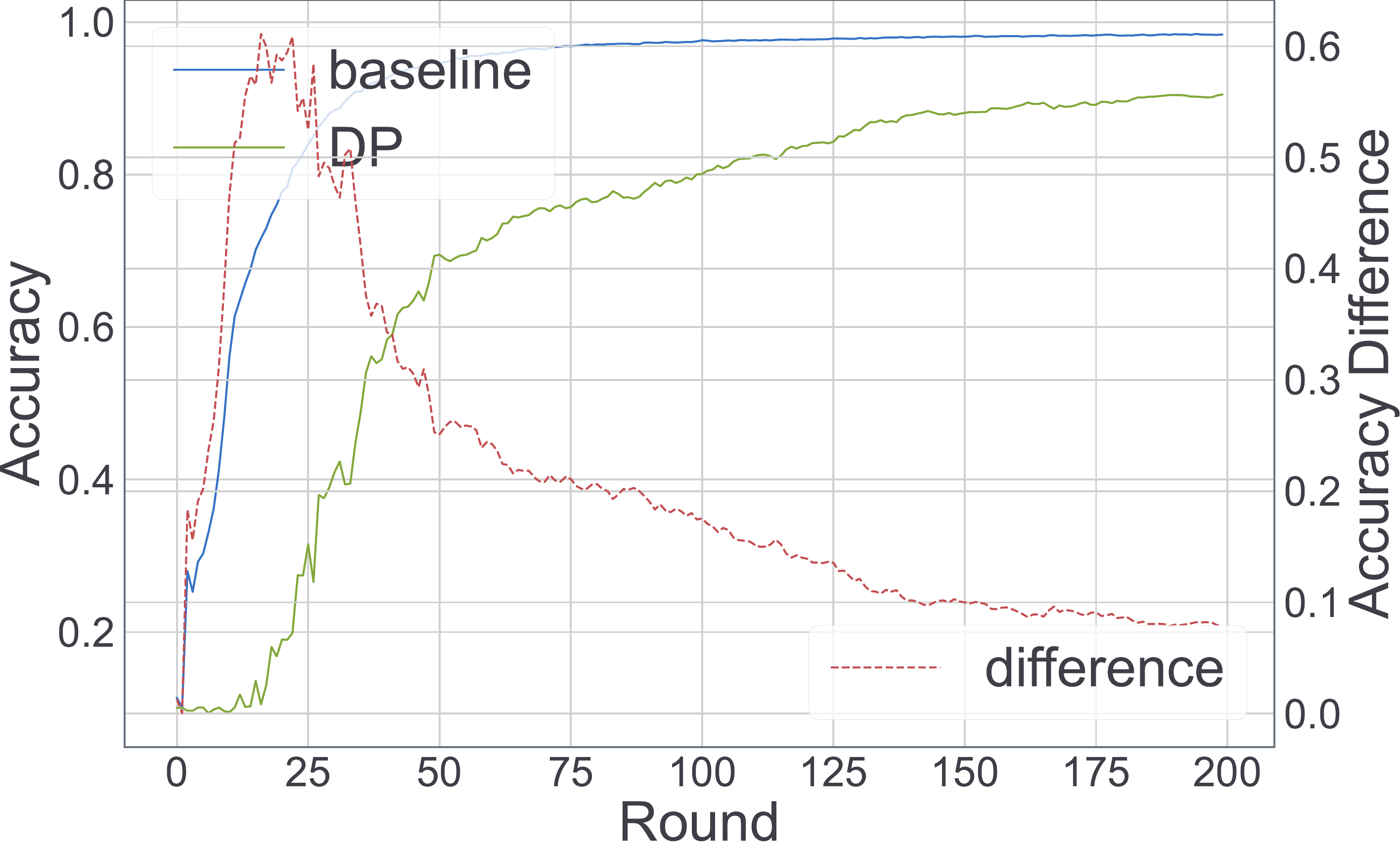}}
    \hfill
    \subfigure[\scriptsize{$\sigma=4$, $q=80/100$}\label{fig: 5-mnist-S4}]{\includegraphics[width=0.3\textwidth]{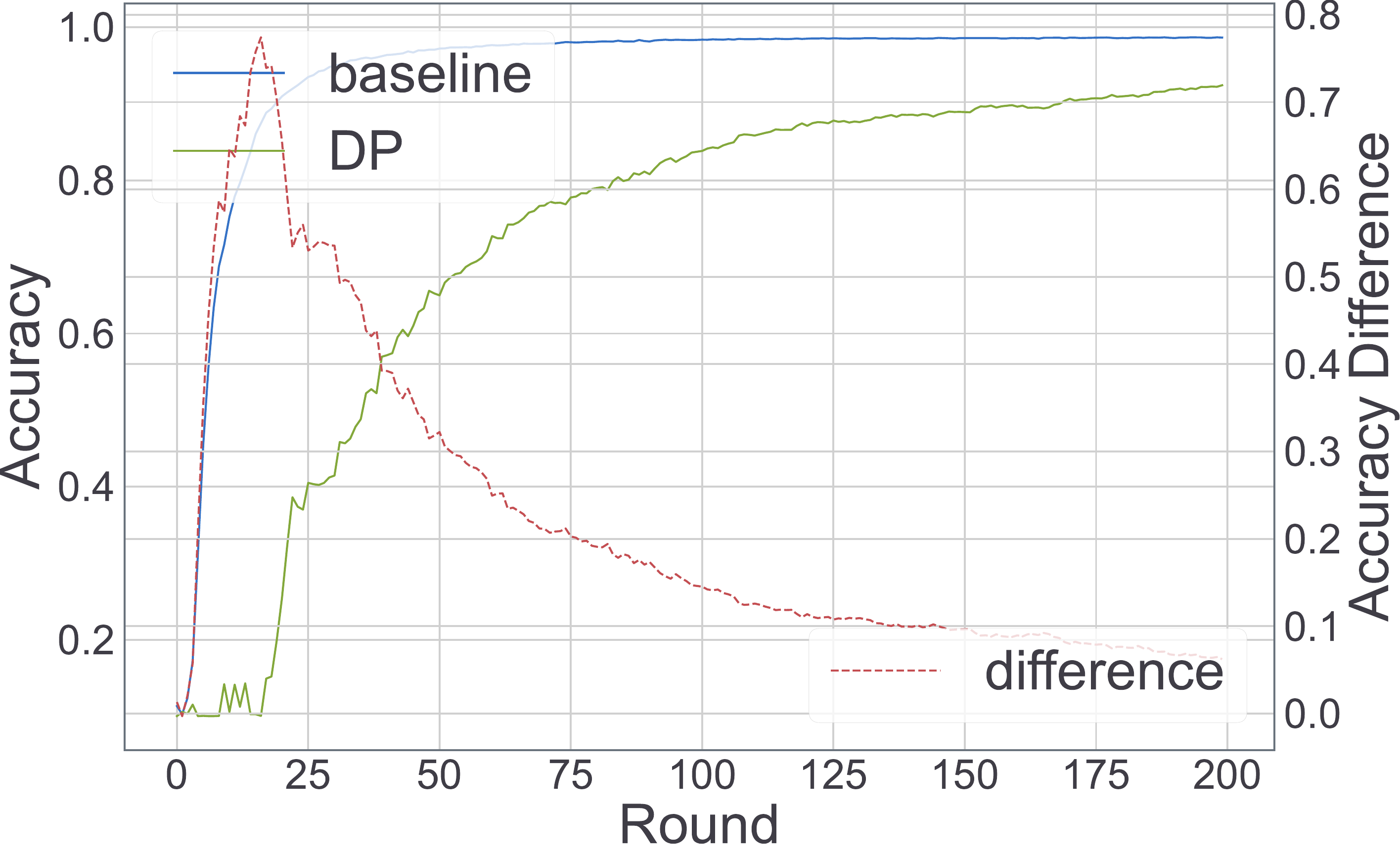}}
    \hfill
    \subfigure[\scriptsize{$\sigma=5$, $q=100/100$}\label{fig: 5-mnist-S5}]{\includegraphics[width=0.3\textwidth]{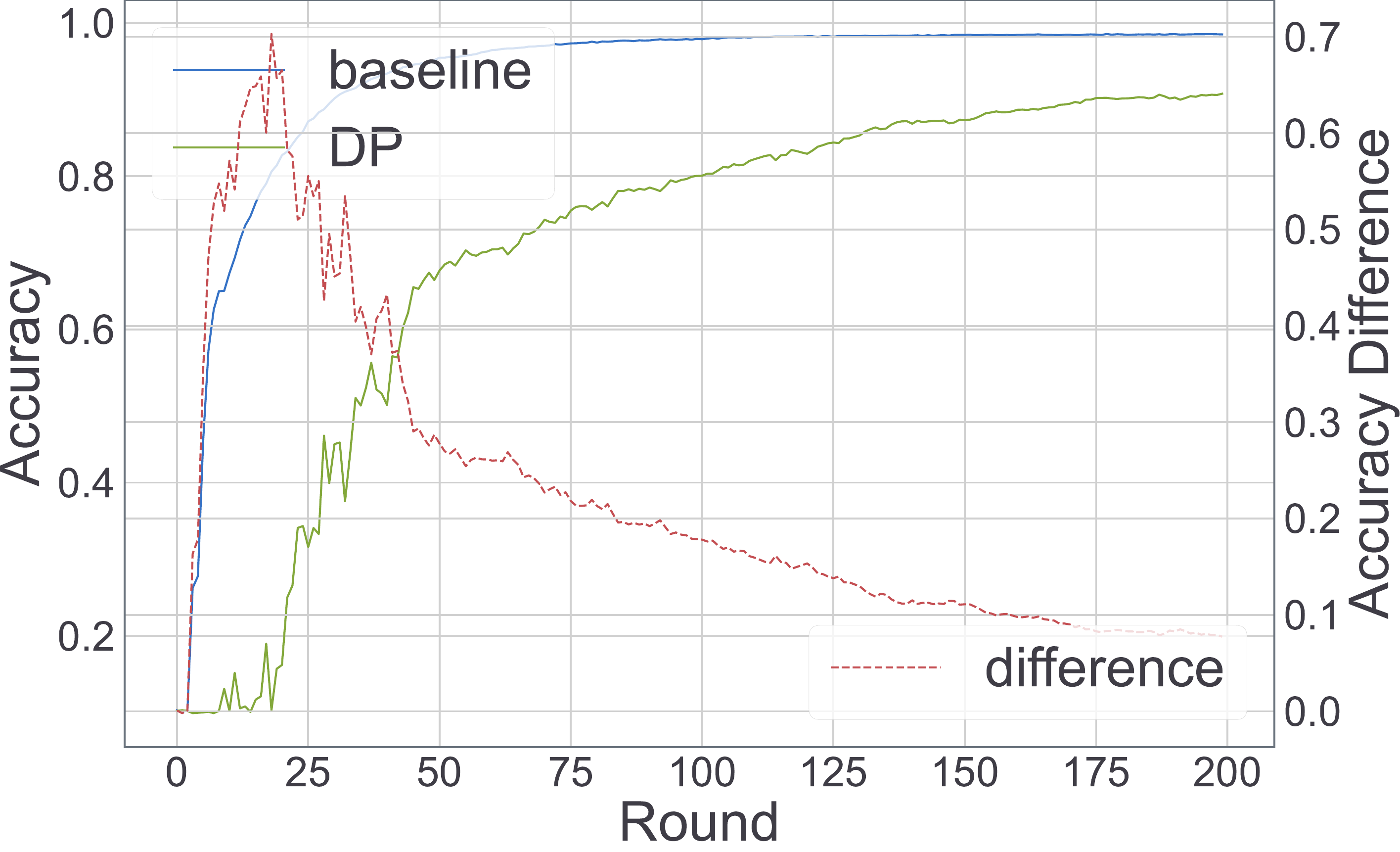}}
    \hfill
    \subfigure[\scriptsize{$\epsilon$ recording}\label{fig: 5-mnist-eps}]{\includegraphics[width=0.3\textwidth]{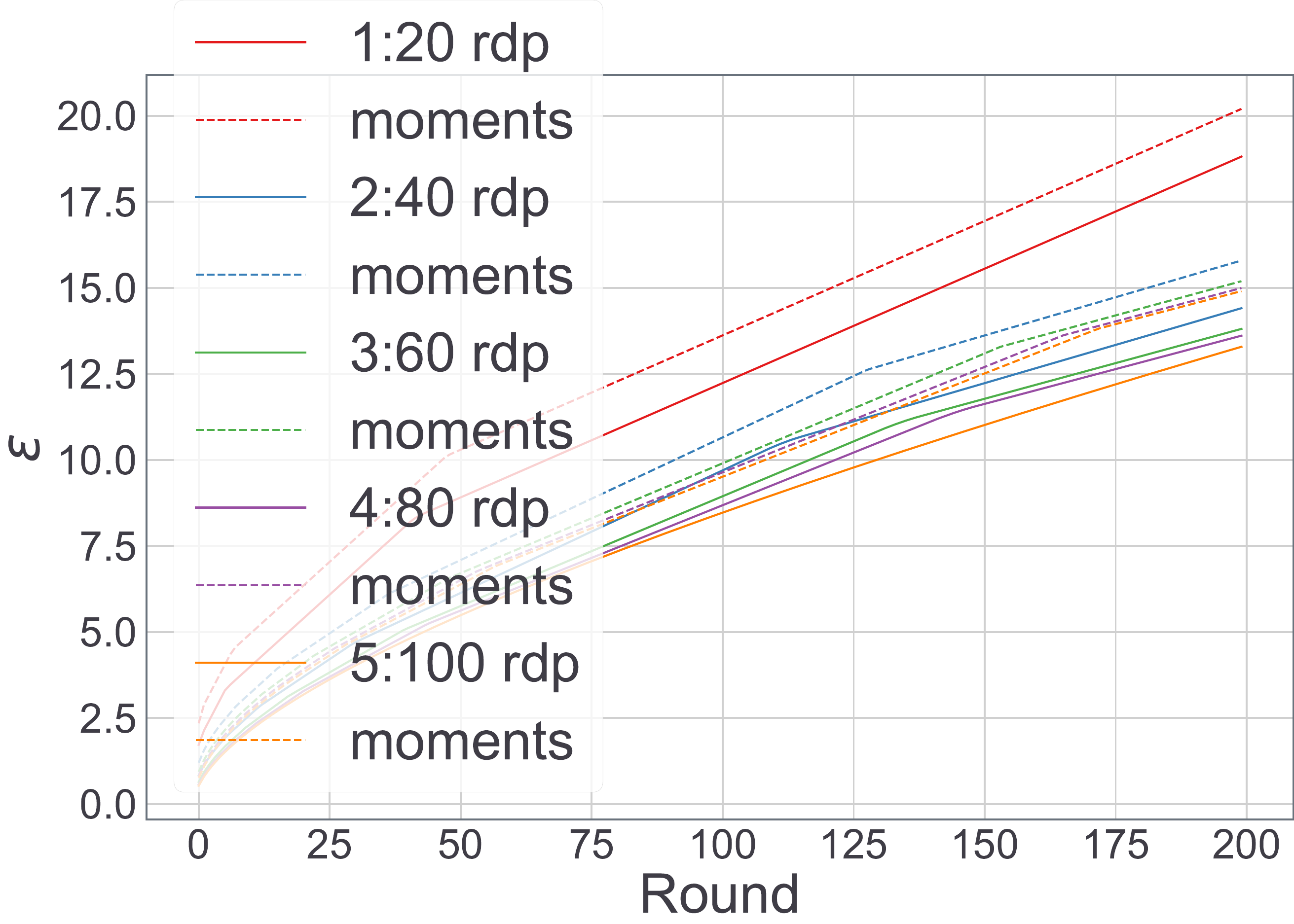}}
    \caption{DP's effect on MNIST}
    \label{fig: 5-mnist-dp}
\end{figure}

The consequence of DP Gaussian noise is tested by MNIST and shown in Fig.~\ref{fig: 5-mnist-dp}. We kept the ratio of noise strength $\sigma$ and the chosen number of clients for each round $n$ constant, $\frac{\sigma}{n}=1/20$. Each experiment used a different subsampling rate $q$ from 0.2 to 1.0. The total number of training rounds extended to 200, as the noise postponed the convergence. Figures nearly exhibit a similar trend. In general, the added noise decreased the final accuracy by 7.0\%. In return, almost in all the experimental settings, $(\epsilon,\delta)$ was better than $(20,10^{-3})$, and the best one can achieve $(13.29,10^{-3})$ estimated by RDP. Solid lines and dash lines in Fig.~\ref{fig: 5-mnist-eps} are $\epsilon$ estimated by Moment's accountant and RDP, respectively, where RDP always provides a lower $\epsilon$'s boundary. The FLVoogd framework provides an adaptive supervisor for $\epsilon$, which the server can customize, so the server can stop the training or adjust the sampling ratio and the amount of adding noise in time once $\epsilon$ is undesirable.

\subsection{Non-IID Interference}
Experiments will test $Deg_{nIID}$ from $0.2$ to $0.7$. The number of clients per round is reduced from 40 to 20 since DBSCAN is unstable with noisy points where the non-IID noise may connect clusters. The defense may collapse if too many clients are selected per round under the non-IID setting.

\begin{figure}[htb]
    \centering
    \subfigure[\scriptsize{A1: PMR=9/20}\label{fig: 5-noniid-A1}]{\includegraphics[width=0.3\textwidth]{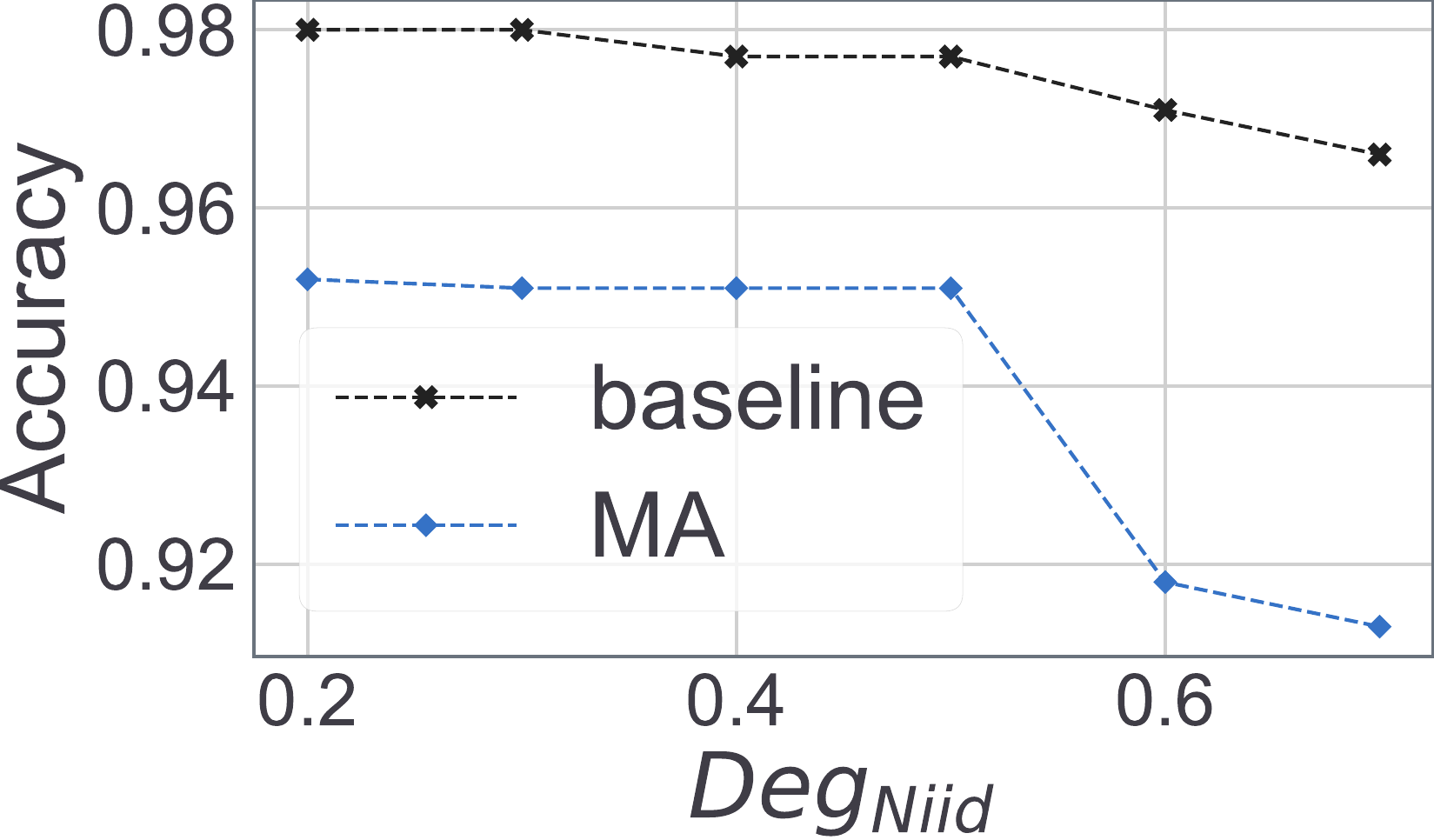}}
    \hfill
    \subfigure[\scriptsize{A2: PMR=9/20}\label{fig: 5-noniid-A2}]{\includegraphics[width=0.3\textwidth]{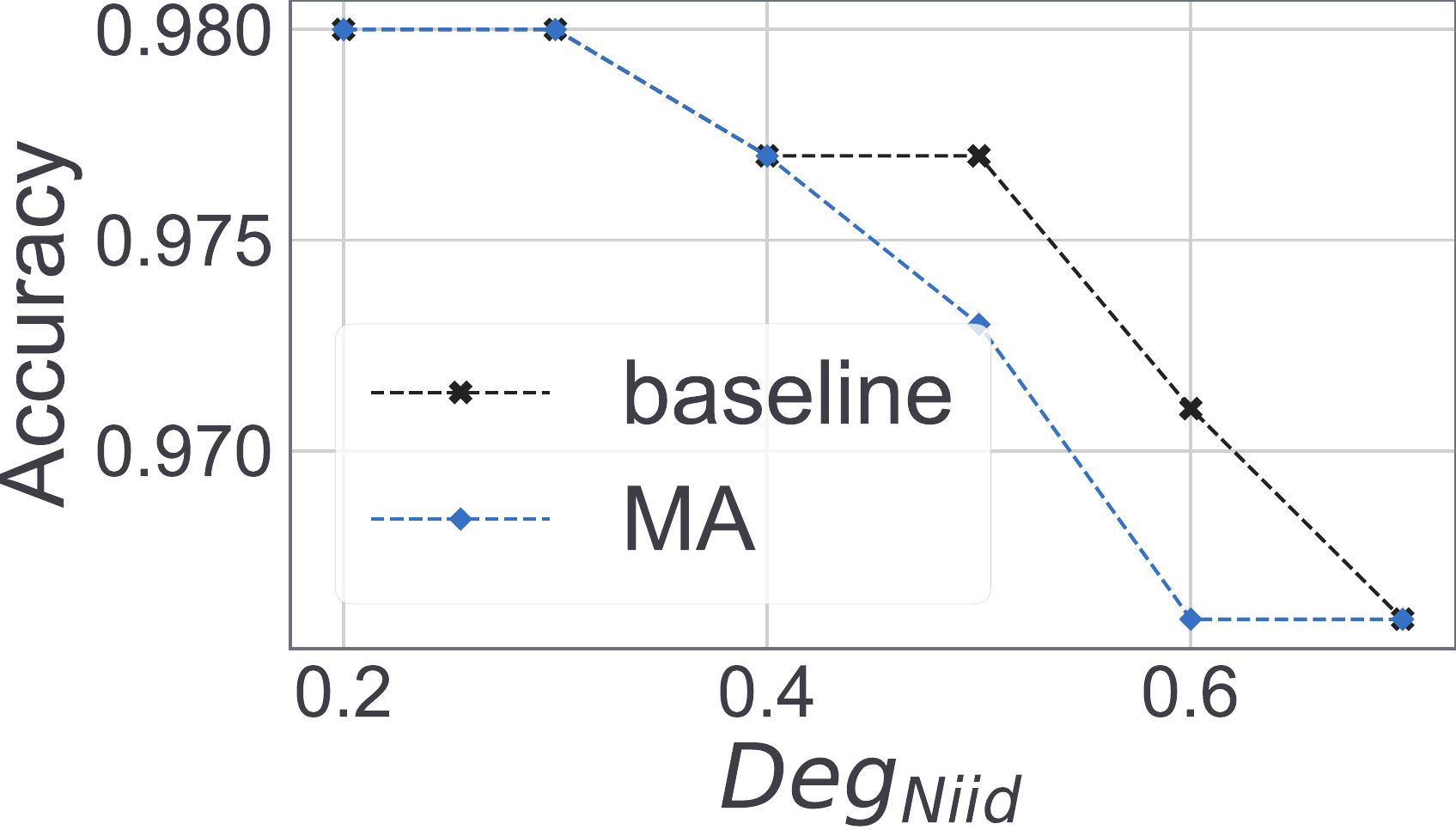}}
    \hfill
    \subfigure[\scriptsize{A3: PMR=9/20}\label{fig: 5-noniid-A3}]{\includegraphics[width=0.3\textwidth]{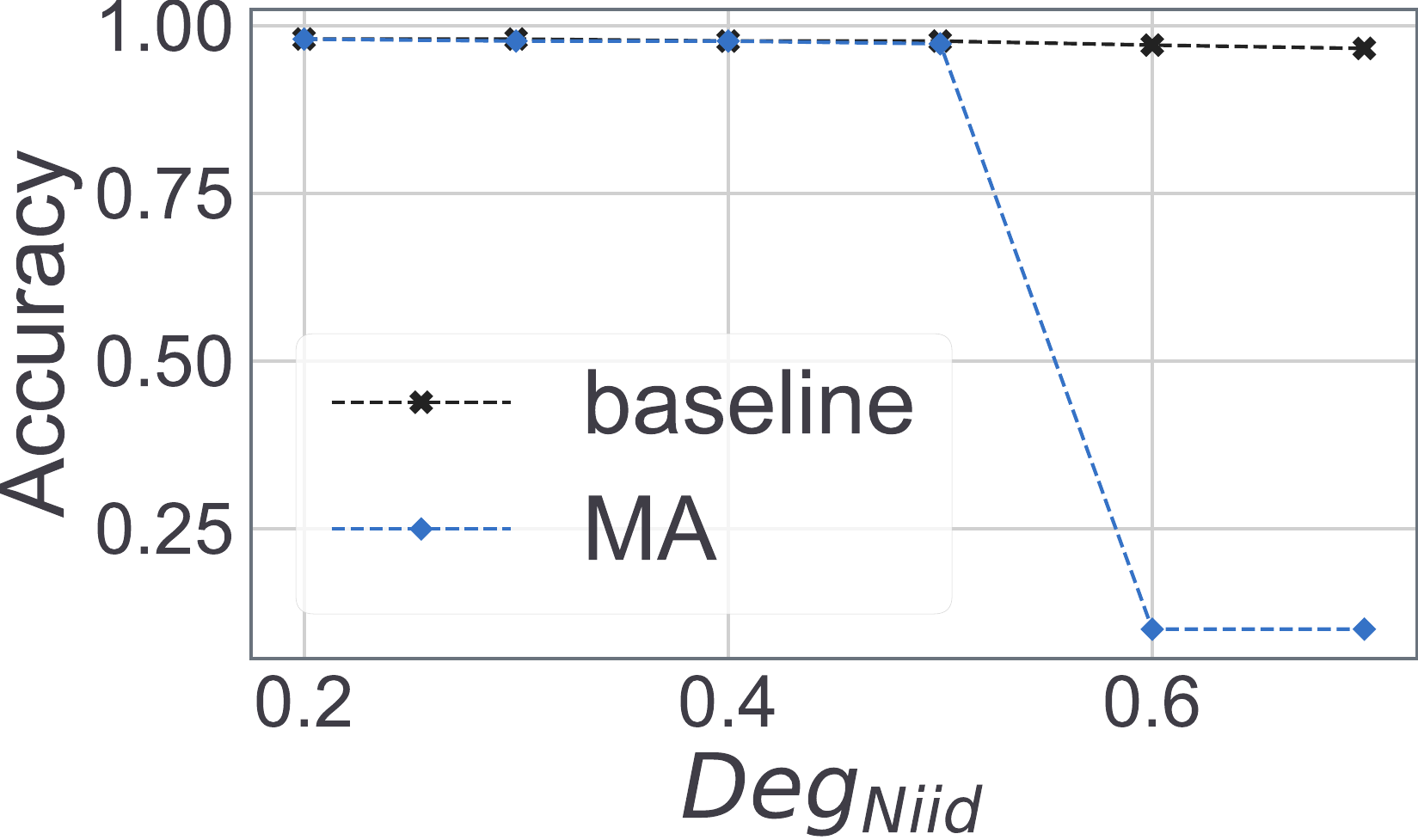}}
    \hfill
    \subfigure[\scriptsize{A4: PMR=9/20, PDR=1.0}\label{fig: 5-noniid-A4}]{\includegraphics[width=0.3\textwidth]{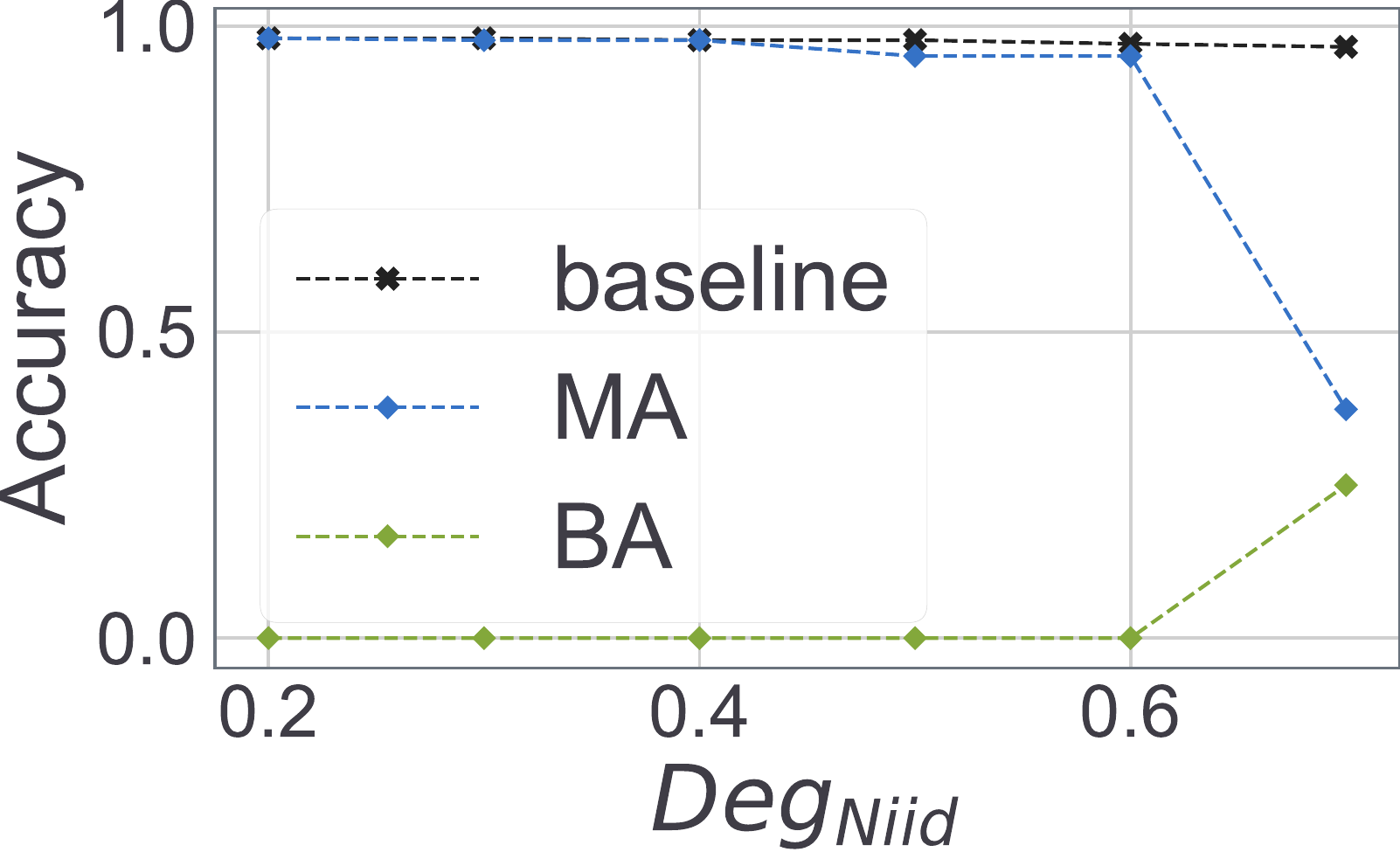}}
    \hfill
    \subfigure[\scriptsize{A5: PMR=9/20, PDR=0.2}\label{fig: 5-noniid-A5}]{\includegraphics[width=0.3\textwidth]{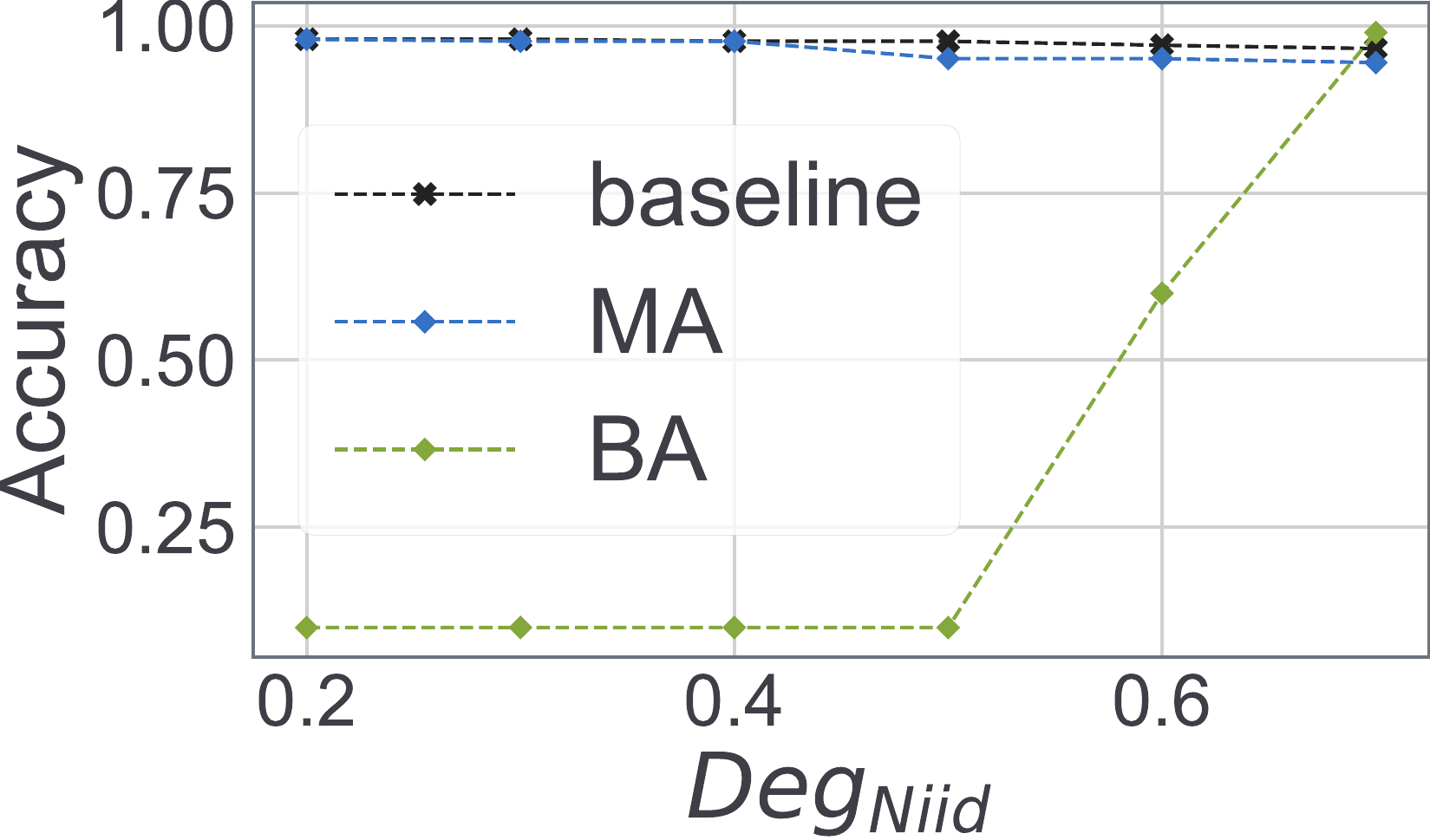}}
    \hfill
    \subfigure[\scriptsize{A6: PMR=9/20, PDR=0.33}\label{fig: 5-noniid-A6}]{\includegraphics[width=0.3\textwidth]{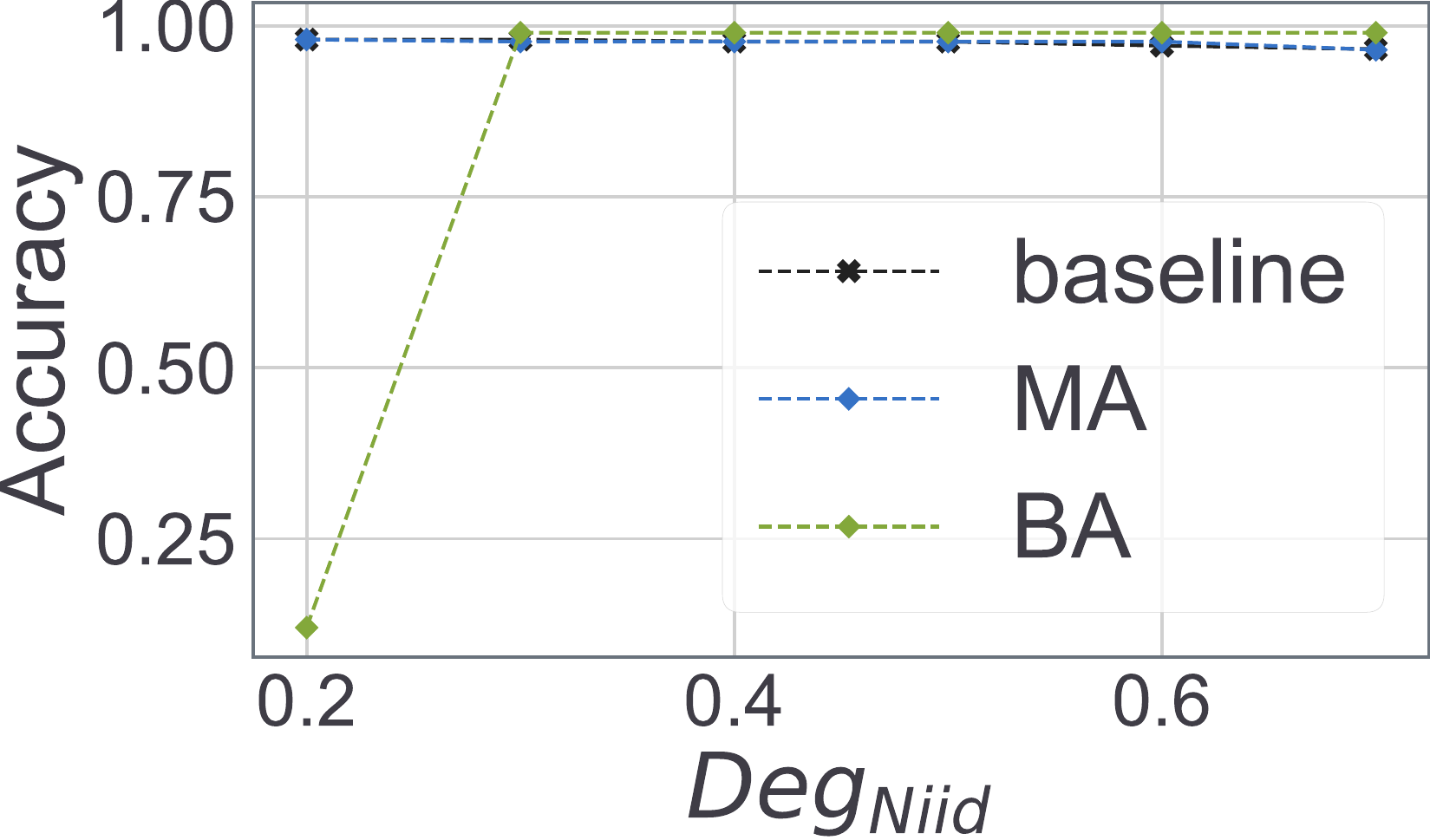}}
    \caption{Non-IID effect on MNIST}
    \label{fig: 5-mnist-noniid}
\end{figure}

The defensive effect under the non-IID setting has been tested on the MNIST dataset, shown in Fig.~\ref{fig: 5-mnist-noniid}. In general, FLVoogd cannot ultimately tolerate that clients hold extreme non-IID samples. In Fig.~\ref{fig: 5-noniid-A1}, the filter rejects all these uploads until the model converges. After convergence, the filter hardly distinguishes between the malicious and model uploads because both appear somewhat random behaviors, resulting in fluctuations in the convergence state. However, the accuracy is still above 90\% since benign uploads again become meaningful once below this threshold, and the filter once more rejects malicious random uploads. Krum attack shows ineffectual regardless of the non-IID degree in Fig.~\ref{fig: 5-noniid-A2}. The filter is so sensitive to the direction of uploads that uploads with reverse direction can scarcely pass the filtering. In Fig.~\ref{fig: 5-noniid-A3}, the filter relinquishes its duty after the non-IID degree is more extensive than 0.5. Compared to A1, A3 is a more advanced scheme, which chooses the randomness adaptively, so the attacking effect is more significant. In Fig.~\ref{fig: 5-noniid-A4}, random flipping works after the non-IID ratio is higher than 0.6. After $Deg_{nIID}>0.5$, one class completely dominates the dataset. In each local batch iteration, one class contributes over 50\% parameters' update. Consequently, the learning process is tampered with by the flipping of one class intermittently once the non-flipped samples of this class miss the training round. In Fig.~\ref{fig: 5-noniid-A5}, backdoor accuracy cannot be constrained if increasing the non-IID degree to more than 0.5, as the filter cannot discriminate whether the non-IID or the backdoor targets cause the directional difference. This situation similarly happens in Fig.~\ref{fig: 5-noniid-A6} where the result is even worse because the defense collapses when the non-IID ratio is just higher than 0.2. Contradicting A5, where the backdoor targets are still the samples in the dataset, A6 introduces the backdoor targets from another dataset and aims to compromise the weakness of the model prediction. The filter performs ineptly if the model digest cannot reflect normal/abnormal directions. Since the model can never learn those edge cases with true labels, the model cannot provide evidence of deviant behaviors. When the non-IID ratio is lower than 0.3, the filter can detect those edge cases mainly because of the distribution of uploads. However, after non-IID increases, the upload lacks this kind of information. 

\begin{figure}[htb]
    \centering
    \subfigure[\scriptsize{$Deg_{nIID}$=0.3: lowering PDR to 6/20=30\%}\label{fig: 5-noniid-q3-pdr}]{\includegraphics[width=0.3\textwidth]{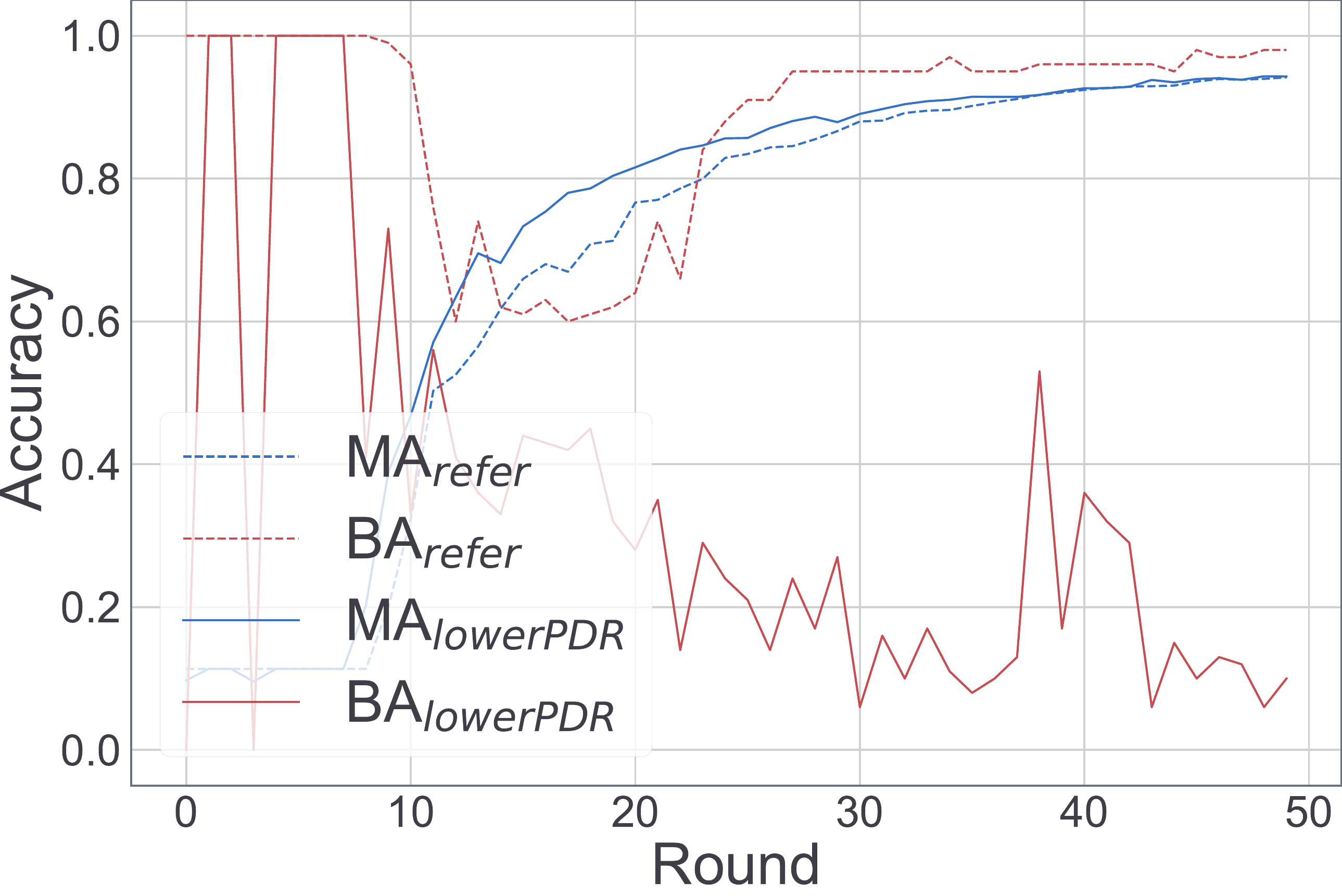}}
    \hfill
    \subfigure[\scriptsize{$Deg_{nIID}$=0.3: pretraining the model to accuracy around 25\%}\label{fig: 5-noniid-q3-pre}]{\includegraphics[width=0.3\textwidth]{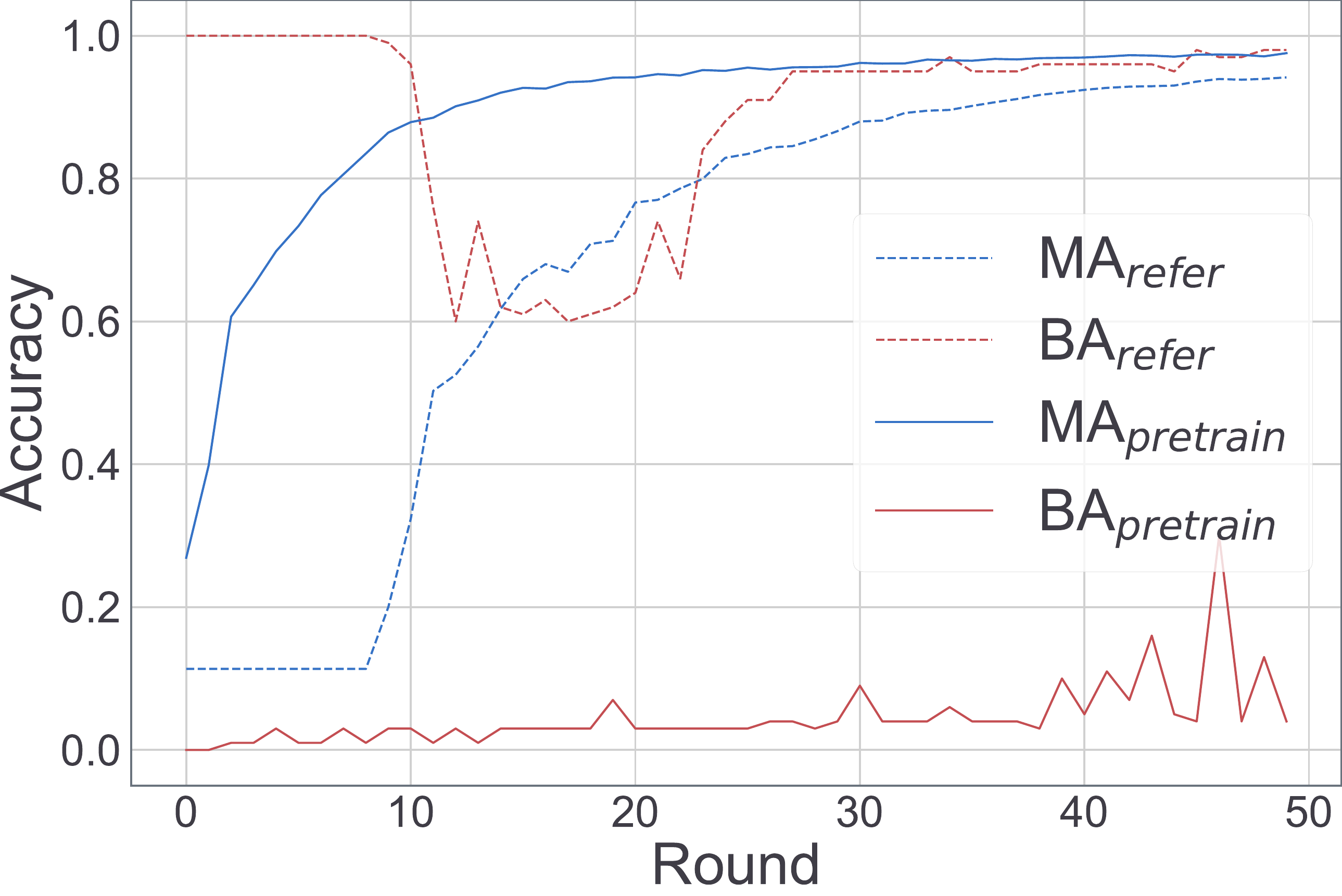}}
    \hfill
    \subfigure[\scriptsize{$Deg_{nIID}$=0.5: pretraining the model to accuracy around 25\%}\label{fig: 5-noniid-q5-pre}]{\includegraphics[width=0.3\textwidth]{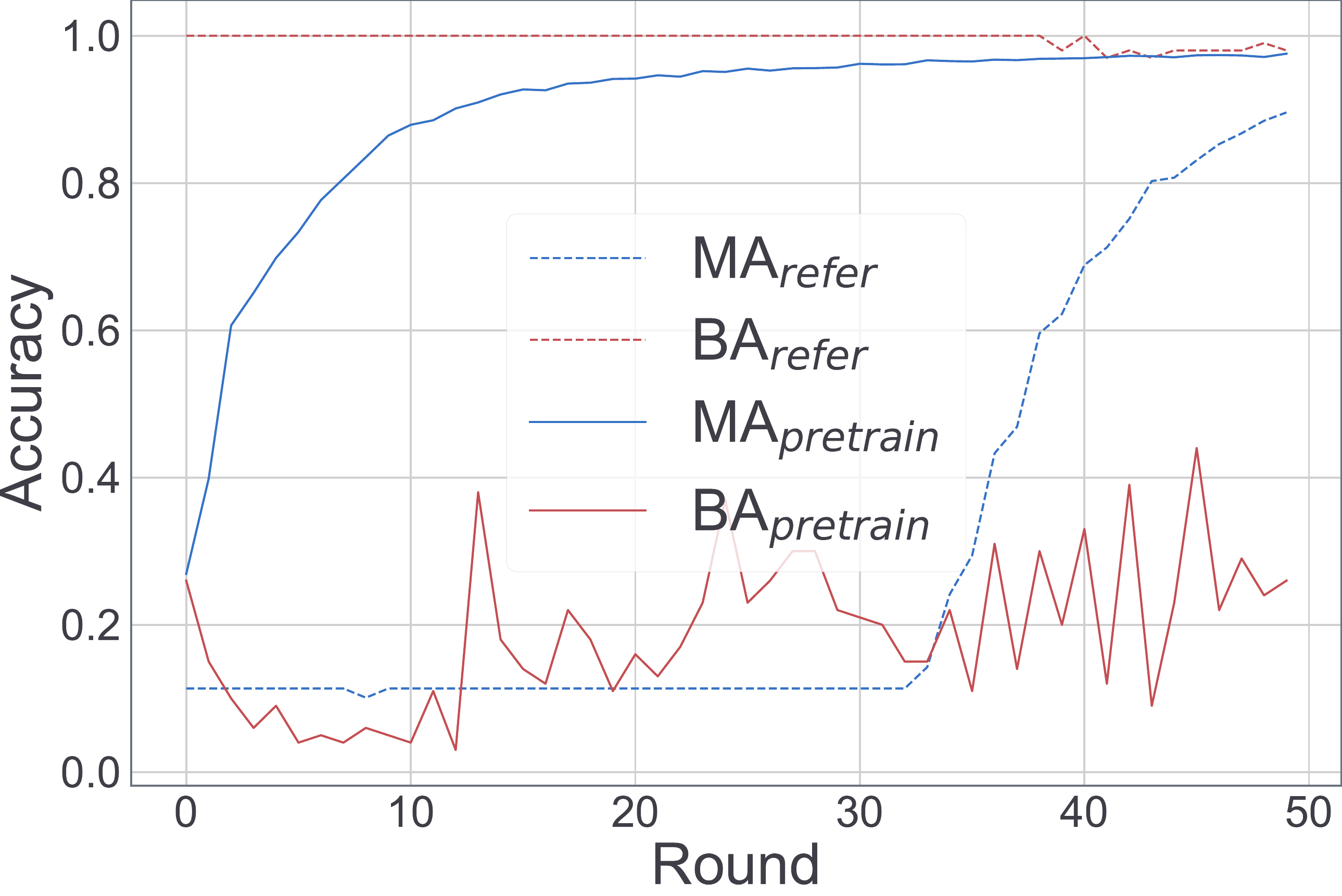}}
    \caption{Pretraining and a lower PDR help the defense}
    \label{fig: 5-mnist-noniid-more}
\end{figure}

Since the FLVoogd performed worse when it suffered from A6's attack, we selected this situation for further study. We wanted to assist FLVoogd somewhat - training the model ahead or decreasing the PMR. As mentioned, pre-training is doable for some application scenarios. In addition, according to~\cite{GoogleFLsurvey}, PMR$\approx50\%$ is a very pessimistic assumption, so we tried to lower it a little bit to see how our framework would react. In Fig.~\ref{fig: 5-noniid-q3-pdr}, PDR is reduced from 45\% to 30\%, and the BA learning curve declines once the model has learned the correct direction from the benign uploads. The poisoning effect is weakened because of the lower PDR. In Fig.~\ref{fig: 5-noniid-q3-pre} and Fig.~\ref{fig: 5-noniid-q5-pre}, the model accuracy is trained approximately to 25\% before the attacks deploy. The updating directions of models become consistent after the pre-train. Thus, the filter can sift those malicious uploads once it first time meets the upload in an abnormal direction. The results also verify that defending against targeted attacks depends on the performance of models on the dataset. If the model can separately recognize the poisoned and normal samples, it can output distinguishable model updates. Then, after the filter captures this variance, the defense effectively works. 

%% file: sections/conclusion.tex
\section{Conclusion}
We introduce FLVoogd, a robust and privacy-preserving federated learning framework that restrains the adverse impact of Byzantine attacks within an acceptable level while maintaining the performance of model predictions on the main task. 
There are two critical differences between our design and prior works. 
Firstly, most procedures are executed under privacy preservation, where operations are doable for mostly popular SMPC protocols. 
Secondly, we provide adaptive adjustments such that the whole process can run automatically. 
Future works could include: merging the transfer learning into the current framework to tackle GAN inference and combine it with other efficiently communicative schemes, e.g., sketch, to reduce the communication bandwidth and enhance the differential privacy.